\documentclass[twocolumn]{aastex631}
\usepackage[bottom]{footmisc}
\usepackage{amsmath}
\usepackage{graphicx}   
\usepackage{bm}
\usepackage{wrapfig}
\usepackage{tcolorbox}
\usepackage[utf8]{inputenc}
\usepackage{longtable}
\usepackage[english]{babel}
\usepackage{natbib}

\usepackage[colorlinks=true]{hyperref}

\begin{document}

\title{Cosmic Collisions I: Optical Morphologies for MIRI-selected galaxies from the PRIMER survey}

%me
\author[0000-0001-5930-0532]{Gregory Troiani}
\affiliation{Department of Physics and Astronomy, University of Kansas, Lawrence, KS 66045, USA}%\email{gtroiani@ku.edu}

%advisor
\author[0000-0002-5537-8110]{Allison Kirkpatrick}
\affiliation{Department of Physics and Astronomy, University of Kansas, Lawrence, KS 66045, USA}

%internal team contributors
%coincidentally alphabetical but also in order of scale of contribution
\author[0000-0002-0717-8122]{Leyna Bajaj}\affiliation{Department of Physics and Astronomy, University of Kansas, Lawrence, KS 66045, USA}\affiliation{Center for Astrophysics Harvard \& Smithsonian, 60 Garden Street, Cambridge, MA, 02138, USA}

\author[0009-0006-9141-4381]{Joseph Havens}
\affiliation{Department of Physics and Astronomy, University of Kansas, Lawrence, KS 66045, USA}		

\author{Jackson Stephens}
\affiliation{Department of Physics and Astronomy, University of Kansas, Lawrence, KS 66045, USA}

%invited authors - comments provided
\author[0000-0002-6610-2048]{Anton M. Koekemoer	}\affiliation{Space Telescope Science Institute, 3700 San Martin Drive, Baltimore, MD 21218, USA}%\email{koekemoer@stsci.edu}			

%invited authors - no comments provided
%alphabetical
\author[0000-0002-1404-5950]{James S. Dunlop}
\affiliation{Institute for Astronomy, University of Edinburgh, Royal Observatory, Edinburgh, EH9 3HJ, UK}
%\email{james.dunlop@ed.ac.uk}

\author[0000-0003-3820-2823]{Adriano Fontana}\affiliation{INAF - Osservatorio Astronomico di Roma, Via Frascati 33, I-00078 Monteporzio Catone, Italy}%\email{adriano.fontana@inaf.it}	
														
\author[0000-0001-9440-8872]{Norman Grogin}\affiliation{Space Telescope Science Institute, 3700 San Martin Drive, Baltimore, MD 21218, USA}%\email{nagrogin@stsci.edu}

\author[0000-0003-4528-5639]{Pablo G. Pérez-González}\affiliation{Centro de Astrobiolog\'{\i}a (CAB/CSIC-INTA), Ctra. de Ajalvir km 4, Torrej\'on de Ardoz, E-28850, Madrid, Spain}%\email{pgperez@cab.inta-csic.es}

\begin{abstract}
We present the first results of visual morphology classifications from the Cosmic Collisions project, comprising 14,657 galaxies from the JWST Public Release IMaging for Extragalactic Research (PRIMER) survey. The sample consists of galaxies detected with MIRI F770W and spans the epoch of peak cosmic star formation ($z\sim 1-3$). We produce rest-frame optical and observed-frame mid-IR images and enlisted citizen science volunteers through the Zooniverse platform to classify galaxy morphology, surface features, and interactions. The project accumulated 304,524 classifications from more than 4,700 volunteers. We find that mid-IR selected galaxies have significantly different morphologies relative to optical and near-IR selected galaxies, favoring dust-rich disk galaxies at $z<2$ and compact systems at $z>2$. This inverts findings from optical/near-IR selection, which shows galaxies becoming more bulge-dominated and elliptical over cosmic time. In addition, we identify a population of galaxies with disk-like optical morphologies and luminous mid-IR point-source cores that may represent obscured AGN candidates, offering a promising path to identify AGNs based on imaging alone. Comparison with visual classifications from HST imaging demonstrates the impact of JWST’s improved sensitivity and resolution on the identification of galaxy structure and interactions, with systems previously identified by HST as compact, smooth, or featureless showing significant extended and internal structure in JWST. These results provide a foundation for future investigations into the connection between morphology, dust structure, star formation, and AGN activity. 
\end{abstract}


\section{Introduction}\label{sec:intro}

Measuring change in galaxy structure throughout the history of the universe offers insight into the mechanisms by which galaxies grow and evolve. Galaxy morphology probes the dynamical state and merger history of the galaxy, and the era of deep surveys with the Hubble Space Telescope \citep[HST,][]{bahcall1989hst} brought to light many fundamental discoveries in galaxy evolution. For example, galaxies in the nearby universe tend to be smooth and bulge-dominated, with disks becoming more prevalent around 7 Gyr ago \citep[e.g.][]{conselice2007, Buitrago_2012}, and peculiar and compact systems becoming more dominant earlier in the universe \citep[e.g.][]{conselice2000_assymetry, buitrago2008_compact, huertas2016mass}. The merger rate has been measured to similar lookback times, and found to increase both with increasing stellar mass and lookback time \citep[e.g.][]{conselice2008_merger, Lotz_2011, Kartaltepe_2012}. This supports the hierarchical view of galaxy growth, in which most stellar mass is assembled through a series of major and minor mergers \citep{white1978core, white1991galaxy, kauffmann1999}. Yet open questions remain, in particular surrounding how and when stellar disks and bulge growth first arise, where in galaxies stars form first and where star formation is quenched, and how and when star formation becomes coupled with morphology, as it is in the local universe \citep{conselice2014evolution, forster2020star}.

There are observational limits to our ability to distinguish the shape and structure of distant objects. For example, cosmological redshift %(traditionally denoted as $z$) 
means optical surveys no longer probe the broad stellar population at $z>1$, resulting in peculiar or clumpy measurements of the restframe UV \citep[e.g.][]{abraham1996morphologies} %I know theres only one here but that's how jeyhan did it for this statement so i'm taking that as permission for this to be okay
which are incomparable to local optical measurements. In addition, the substructure of distant galaxies is difficult to resolve, with angular size scales reaching a minimum at $z\sim1.0-1.5$ and staying relatively small as redshift increases \citep{weinberg1972, weedman1986, peebles1993}, while empirical studies \citep[e.g.][]{L_PEZ_CORREDOIRA_2010, Raikov_2025} show galaxies tend to decrease in angular size with redshift. Likewise, surface brightness fades rapidly with redshift, dropping the faint features associated with galaxy disturbances \citep{johnston2008_lsb, cooper2010_lsb} below the level of detectability \citep{Lotz2004}. %I dug for awhile to try and find a detectability study for this. Finally I found my way to erini's paper and she says basically this exact sentence and cites the gini-m20 paper, so i'm just trusting her judgment and doing that.

%NOTE - I pulled this paragraph from the serendipitous science section - not sure if it has a place here, or if I should just scrap it.

%There are many ways to identify active galaxies at varying stages of their duty cycle. The most physically direct probes of black hole activity require observationally expensive or targeted spectroscopy (e.g. high ionization or broad lines), radio, X-Ray, or time series observations \citep{padovani2017agn}. In large surveys without specific targets, AGNs can be identified through color selection in the optical (e.g \cite{Richards_2004}) and infrared (e.g. \cite{miricolors}), but these methods are subject to contamination from hot stars (CITATION), silicate dust (CITATION), and PAHs at some redshifts (CITATION). Attempts to discern AGN populations through SED fitting suffer from similar issues. However, while silicate dust does tend to exhibit more compact structure than the stellar components of galaxies \cite{rujopakarn2016}, these contaminants are unlikely to be concentrated into a luminous point source in the core (CITATIONS?). 

The launch of the James Webb Space Telescope \citep[JWST, ][]{gardner2006jameswebb} pushed the limits of sensitivity, resolution, and redshift window, giving us access to a new population of faint, distant galaxies. The Near-Infrared Camera \citep[NIRCam, ][]{nircam} has high angular resolution ($\sim$0.1") out to 5$\mu$m, allowing for well-resolved rest-frame optical measurements at $z\le5.6$. Already, JWST observations have revolutionized our understanding of morphological development, with fully-evolved disks and bulge-dominated systems emerging much earlier in the universe than previously thought possible \citep[e.g.][]{ceersKartaltepe2023, huertas-company2024_jwstmorpho}. 

%The state of the stellar populations probed by optical morphology is, however, not the only structural parameter of interest. Galaxies also contain reservoirs of gas and dust. These reservoirs have long been understood to be highly collisional and turbulent \citep[e.g.][]{spitzer1951_ism, lin1964_spiral, shu1973_shocks}, leading to vastly different dynamical behavior and necessitating independent investigations of structure. The morphology of cold molecular gas, through CO observations in nearby galaxies, has been classified and found to trace spiral structure \citep{helfer2003_codata, leroy2009heracles}, as well as bar structure, even in cases where bars are not identified in the optical \citep{stuber2023}. The thermal emission of cool, large grain dust, which broadly traces the same molecular gas supply as CO \citep{boulanger1996_dustgas}, has been resolved and described morphologically for high-$z$ populations at sub-millimeter wavelengths \citep[e.g.][]{simpson2015_alma, hodge2019_alma, gullberg2019_alma}. These dust structures tend to be compact compared to the optical component, and suggest that bars are an essential mechanism of stellar bulge growth in the early universe. 

%While these observations trace the gas supply which fuels star formation, they do not necessarily describe the structure or location of star formation itself. 
Because the majority of star formation occurred at $1<z<3$ \citep[generally referred to as `cosmic noon',][]{cosmicNoon}, but most star formation in this epoch is hidden under dust \citep{magnelli2011_dustobscuredSFR, moutard2020_dustobscuredsfr}, morphologies of dust-obscured star formation are highly desirable to understand galaxy mass assembly. Continuum emission from very cold dust, which can trace obscured star formation, has been done \citep{rujopakarn2016, mitsuhashi2024}, and shows that star formation is extended and largely co-spatial with the stellar population, with compact dust structure requiring extreme star formation. However, this is relatively observationally expensive, making this method impractical with large samples. 

Observations in the mid-infrared (mid-IR) probe emission from polycyclic aromatic hydrocarbons (PAHs).%, which are an essential component of dust, alongside carbonaceous and silicate grains \citep{Draine_2001, Li_2001_dust, draine2007_dust}. 
Because PAHs emit when excited by UV photons associated with young stars, %but emit in wavelengths that pass through dust, 
they serve as an excellent tracer for star formation which is otherwise obscured \citep{Peeters_2004_pahSFR, Brandl_2006_pahSFR, Shipley2016}. %PAH emission at 3.3$\mu$m has been described \citep{Buta_2010_spitzermorph} with the Spitzer Space Telescope \citep{werner2004_spitzer}, and largely traces optical morphology, but this was performed at redshift of effectively zero. 
The Mid-Infrared Instrument \citep[MIRI, ][]{miri} aboard JWST offers sub-arcsecond resolution between $5.6 - 25.5\mu$m, allowing PAH emission to be resolved in the epoch of peak star formation for the first time. The relationship between optical structural parameters and JWST/MIRI sizes and Sérsic profiles was investigated in \cite{Lyu_2025}, which found evidence of secular inside-out bulge growth and quenching, but largely, little work has been done on the resolved structure of obscured star formation in these distant, dusty galaxies. 

%In addition to being able to resolve IR structure for the first time beyond the local universe, 
Additionally, MIRI is well-suited to detecting the reprocessed dust torus emission of obscured active galactic nuclei (AGNs) \citep{rowanrobinson1989_torus, kirkpatrick2012_IRSEDs, Kirkpatrick_2015_IRSEDs, padovani2017agn}. AGNs are subject to a number of open questions. Notably, as much of their growth is 
expected to be hidden behind dust with increasing lookback time \citep{Ananna_2019} and difficult to find \citep{kirkpatrick2023}, the mechanisms by which they grow remain the subject of debate. In particular, the role galaxy interactions play in igniting AGN growth is unclear. Galaxy collisions provide one mechanism for gas to lose angular momentum and infall, triggering accretion onto central supermassive black holes \citep{sanders1988_bhmerger, dimatteo2005_merger, hopkins}. Observations of the connection between mergers and AGNs are mixed. For example, optical/IR selected AGNs show a correlation with mergers at z$<$0.6 \citep{gao2020}, but when the most luminous/massive members of this population are examined in the X-ray, this correlation disappears \citep{villforth2017}. The picture in the distant universe is similarly unclear. X-ray selected AGNs at 1.5$<$z$<$2.5 show no preference for residing in mergers \citep{kocevski2012}, but IR-selected AGNs at comparable redshifts do show a preference as obscuration increases \citep{kocevski2015}. %Connections may be a selection-dependent, since black holes in mergers tend to be obscured \citep{Ricci2017_obscuredBHmergers}. Nonetheless, further investigations into the types of galaxies which host AGNs are needed. 

%flesh out science questions here
% %would this work better as an itemized list? I wrote it that way but it takes up a lot of space.

%add stuff here about
% color selection as proxy for morphology
% parametric and nonparametric measurements
%To investigate these questions, we must first build a sample of optical morphology classifications against which to compare other measurements. 
There are a variety of methods which can extract morphology information from an image, and all have their advantages and drawbacks. Nonparametric measurements \citep[e.g.][]{Bershady2000, Abraham_2003, CAS, Lotz2004} may be rapidly computed and used as a proxy for morphology. However, these metrics are highly systematically sensitive to depth, resolution, and size scale \citep{statmorph-lsst}, all of which vary strongly with redshift. In addition, while various combinations of metrics can effectively select mergers when used properly, different metrics are sensitive to different stages of the merger process \citep{freeman2013, Wen_2014, pawlik2016, Peth2016}. While machine learning tools for galaxy classification grow ever more sophisticated, and reliable/flexible machine learning models exist for galaxy morphology specifically \citep{zoobot}, classification via machine learning requires abundant training data which is comparable to the objects to be classified. Projects using visual classifications of JWST images exist \citep{ceersKartaltepe2023, zoojwst_1, zoojwst_2}, but there is no publicly available catalog of classifications at the resolution and redshift probed by JWST to use as training data. %Therefore, the only path forward is producing our own set of visual classifications.

Visual classification has been a standard method of galaxy classification for nearly a century \citep{hubble1936} and for most of that time it was performed by hand by small teams of professional astronomers \citep[e.g.][]{sandage1961, devaucouleurs1991}. 
In the era of all-sky surveys with thousands to millions of sources, it has become necessary to crowd-source the visual classification process. This was pioneered by Galaxy Zoo \citep{galaxyzoo}, which demonstrated the scientific validity of galaxy morphology catalogs produced by the general public, with $>90\%$ agreement between classifications from citizen scientists and those produced by professional teams. The success and popularity of Galaxy Zoo has led to the widespread adoption of citizen science initiatives across many fields, including social sciences \citep[e.g.][]{blickhan2019, cartwright2019}, medicine \citep[e.g.][]{spiers2020, quan2023}, and ecology \citep[e.g.][]{rowley2019frogid, black2023}. The Galaxy Zoo project itself has produced visual classifications for $>10^6$ galaxies from the Sloan Digital Sky Survey \citep[SDSS,][]{york2000SDSS} \citep{galaxyzoo1, galaxyZoo2}, HST \citep{willett2017_GZHST, galaxyZooCANDELS}, and many other surveys \citep[for a more comprehensive review, see][]{Masters_2019},
while ongoing projects classify data from JWST \citep{cosmosweb} and the Rubin Observatory \citep{ivezi2019LSST}. This data has been used to produce major results on many topics of interest in galaxy evolution, including the chirality of spiral galaxy rotation \citep{land2008, slosar2009_chiral}, the way star formation and structure depend on environment \citep[e.g.][]{bamford2009, skibba2012, Smethurst_2017}, and more.

In this paper we present the results of our citizen science classification of mid-IR selected galaxies at cosmic noon, which we publicly release. In Section \ref{sec:data}, we describe the PRIMER reductions and sample selection, as well as the creation of our classification images. In Section \ref{sec:method}, we explain our classification system and data pipeline, and in Section \ref{sec:results}, we present our processed classifications and redshift evolution. Section \ref{sec:conclusions} summarizes our findings. Throughout this paper, we assume a standard flat cosmology with $H_0 = 70\ \text{km}\ \text{s}^{-1}\ \text{Mpc}^{-1}$, $\Omega_{m} = 0.3$, and $\Omega_\Lambda = 0.7$.  

\section{Cosmic Collisions Overview}\label{sec:data}
\subsection{Science Motivation}
The goal of the Cosmic Collisions project is to investigate the myriad questions surrounding mid-IR morphology, optical morphology, and mid-IR spectral shape. Specifically: Do obscured AGNs reside preferentially in interacting and recently merged systems, especially at high redshift? Does the structure of dust and obscured star formation visible in MIRI observations trace the stellar population and unobscured star formation probed with NIRCam? Are galaxies with significant mid-IR emission morphologically distinct from the broader galaxy population, and, if so, how do these differences evolve with cosmic time? This work focuses on addressing the third question, while laying the groundwork for future efforts towards the other two.

Cosmic Collisions currently consists of more than 10,000 visually classified JWST images. The catalog of classifications is publicly available in the supplemental material for
this paper.

\subsection{PRIMER and Sample Selection}

The data for this work comes from the Public Release IMaging for Extragalactic Research (PRIMER) survey. PRIMER is a Cycle 1 JWST treasury program (PID: JWST-GO-1837, PI J.S. Dunlop) which covers portions of the COSMOS 
(10h\,00m\,31.20s, 02$^\circ$\,21$^\prime$\,36.00$^{\prime\prime}$) and UDS (02h\,17m\,26.40s, $-05^\circ$\,12$^\prime$\,36.00$^{\prime\prime}$) fields. It observes with eight NIRCam filters spanning $0.9 - 4.4\mu$m and 2 MIRI filters at 7.7$\mu$m and 18$\mu$m. PRIMER is one of the largest area JWST surveys, covering 378 arcmin$^{2}$. Imaging was reduced with \texttt{GRIZLI} \citep{grizli} and all NIRCam images were scaled to 0.04 ''/pix. The PRIMER photometry catalog was built using \texttt{Aperpy}\footnote{https://github.com/astrowhit/aperpy} on a combined F277W+F356W+F444W detection image, with settings matching those used in \cite{weaver2024}. This catalog was cross-matched with photometry from the HST CANDELS survey \citep{CANDELS}. Redshifts and stellar parameters (including stellar mass, star formation rate (SFR), and V-band model luminosity) were determined with \texttt{EAZY}\footnote{https://github.com/gbrammer/eazy-py} \citep{EAZY} using all available HST and NIRCam bands with the \textit{SFHz\_CORR} templates\footnote{https://github.com/gbrammer/eazy-photoz/tree/master/templates/sfhz}, which includes a high redshift emission line template from \cite{highztemplate}, as well as a high redshift AGN/LRD template from \cite{agntemplate}. This produced a catalog of 296,134 objects.

The PRIMER MIRI F770W and F1800W imaging mosaics were processed as described in \cite{perezgonzales2024}. In the COSMOS field, the PRIMER mosaic for F770W was also reduced in conjunction with observations with the COSMOS-Web Survey \citep{cosmosweb}, another Cycle 1 JWST treasury program. We constructed a mid-IR catalog by using \texttt{photutils} \citep{photutils} on the F770W images with the followings settings: the connected pixels threshold was set to 9; the number of deblend levels was set to 32; the deblend contrast was set to 0.005; and the detection threshold was set at 1$\sigma$. We then performed photometry on elliptical apertures generated by \texttt{photutils}, scaled up by a factor of three, in both F770W and F1800W, with both background and Poisson error included. We selected this aperture size based on visual inspection.
Then, we imposed a cut of SNR$>$3 in F770W before matching with the existing NIRCam catalog, with a spatial matching tolerance of 0.29$^{\prime\prime}$, which is the full-width at half-maximum (FWHM) of the F770W point-spread function (PSF). This produced a photometric catalog of 16,952 NIRCam and MIRI detected objects, with 9334 in the COSMOS field and 7618 in the UDS field.

\begin{figure}[ht]
\includegraphics[width=0.5\textwidth]{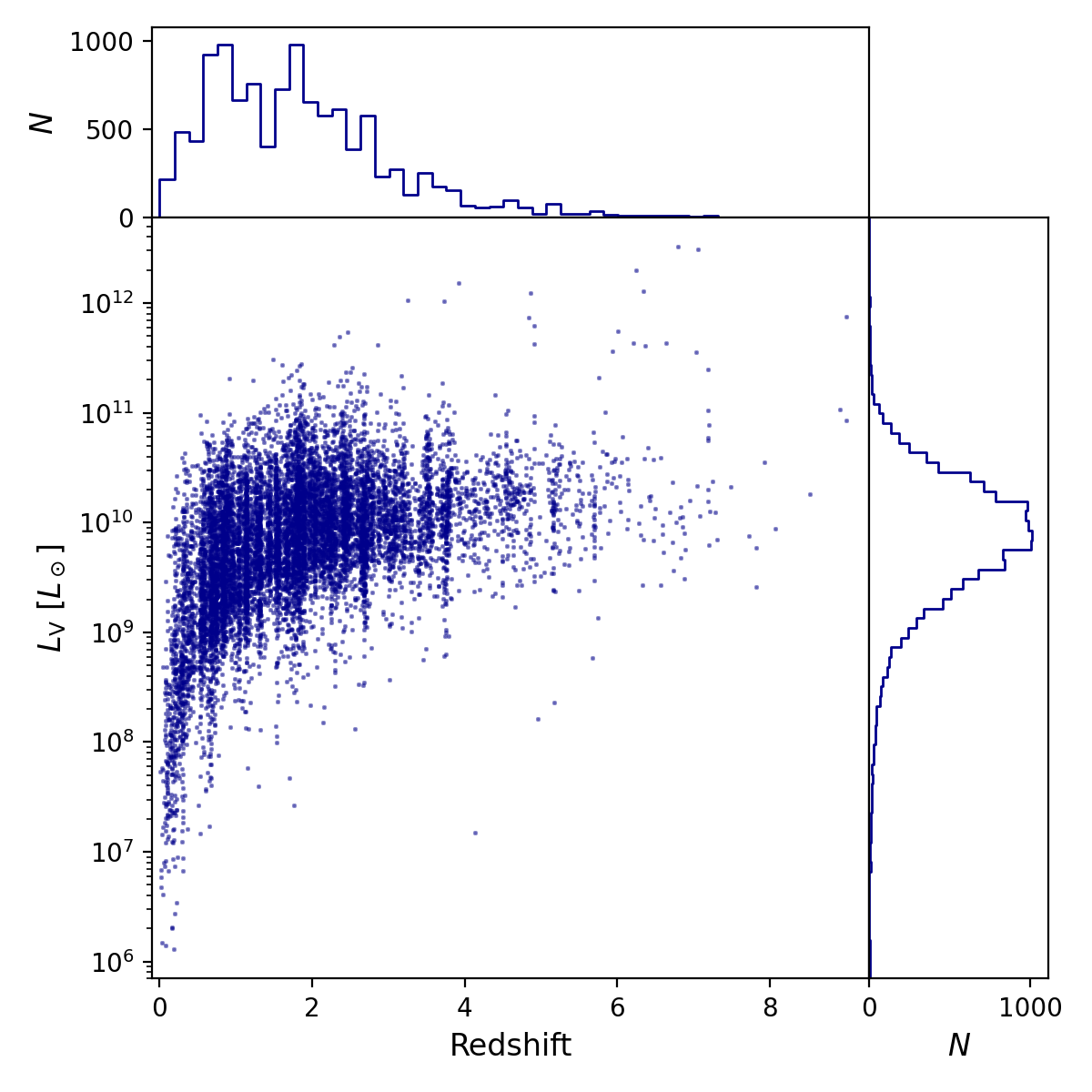}
    \caption{The rest-frame V-band luminosity $L_{\rm V}$ plotted against redshift. The side panels show histograms of each variable. The bulk of our objects are galaxies with $9<\text{log }L_{\rm V}/L_\odot<11$ residing at cosmic noon. 
    \label{fig:LvsZ}}
\end{figure}

For the visual classification dataset, we made further cuts to ensure that a reasonable image could be produced. First, we require acceptable redshifts based on the reduced $\chi^2$ of the \texttt{EAZY} fits. We also require SNR$>$10 in F770W and at least two NIRCam bands. This resulted in the final classification set of 14,657 objects, with 7485 in the COSMOS field and 7172 in the UDS field. The redshift and luminosity distribution of these objects are shown in Figure \ref{fig:LvsZ}.

\section{Citizen Science Classification}\label{sec:method}

\subsection{Image Creation}\label{subsec:images}
\begin{figure*}[hbt]
\includegraphics[width=1\textwidth,]{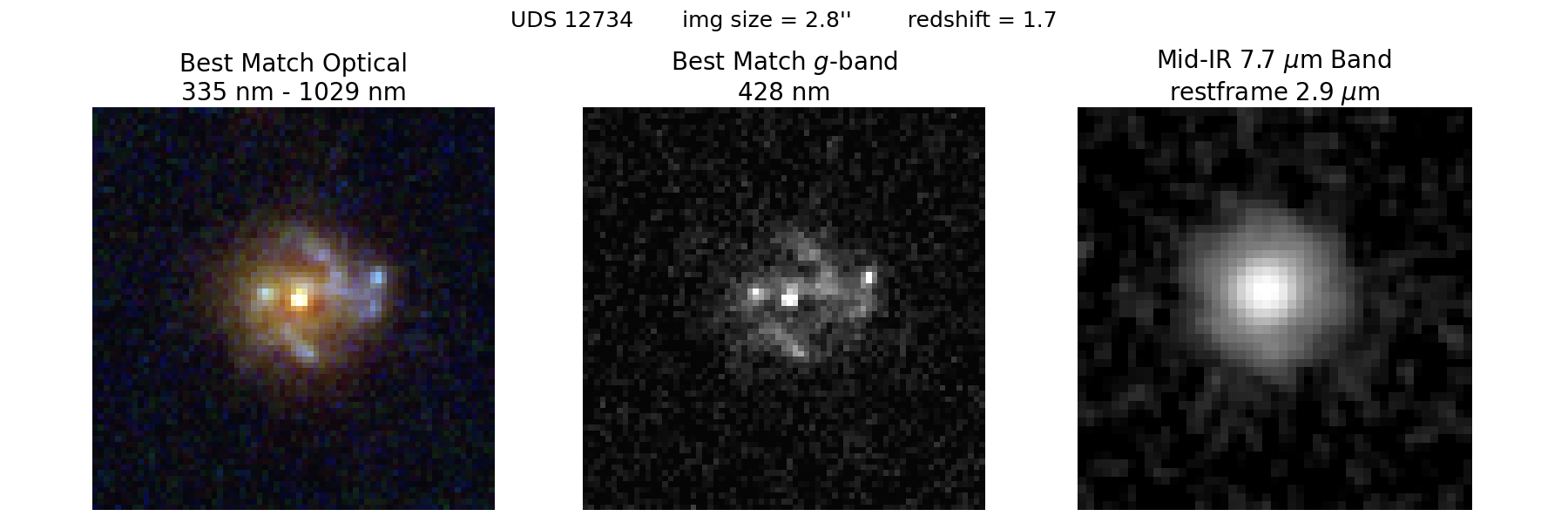}
    \caption{A Zooniverse visual classification image. The leftmost panel shows a color composite created from all available NIRCam bands span the rest-frame optical. The middle panel shows the same object in the closest match to the rest-frame $g$-band, and the right panel shows the object in the F770W filter. The image size, object ID, and redshift are given in the header. The F770W image shows a starkly different morphology, and citizen scientists are given the opportunity to flag such galaxies on the talk pages.
\label{fig:image_example}}
\end{figure*} 

We produced color composite images using a modified version of Trilogy \citep{trilogy}. The color image is comprised of all available NIRCam data which falls into an approximate rest-frame optical spectrum (between 380 and 750 nm). We chose to approximate an optical visualization in order to make our classification images as scientifically homogeneous as possible. The method of selecting NIRCam bands is designed to be flexible in order to ensure that the entire optical spectrum is spanned, when available; as such, a small amount of near-UV and near-IR is often included, as illustrated by the wavelength range in Figure \ref{fig:image_example}. When the redshift is such that no NIRCam bands fall within the rest-frame optical (roughly $z<0.2$ or $z>10.6$), then all available NIRCam bands are displayed in the color image, with the corresponding rest-frame wavelengths displayed. This occurred for 233 low-$z$ objects and only 2 high-$z$ objects. 

We also include a grayscale image of the filter which is the closest match to the rest-frame $g$-band and an F770W image. The former is included to give users a simplified version of the color composite image while also focusing on the location of unobscured star formation. For the 235 objects with no optical information, this panel instead shows a grayscale image of the filter that is closest to the optical (generally F090W for low-$z$ and F444W for high). The latter is included so that users can note any objects with significant differences in shape/structure between the rest-frame optical and the observed mid-IR. Figure \ref{fig:image_example} shows an example of one classification image containing these three panels. 

The image size is scaled to have a side length of 10 times the semi-major axis (as determined by \texttt{Aperpy}) of the central source. We selected this scaling based on visual inspection as consistently zoomed in enough to visualize detailed structure, while showing enough of the galaxy's surroundings to see the nearest neighbor. We also prevented images from scaling larger than 10 arcseconds to prevent catastrophically large images due to contamination by diffraction spikes. This upper limit affected 318 objects in the dataset.

\subsection{Zooniverse Project}\label{subsec:zooniverse}

\begin{figure}[ht]
\includegraphics[width=0.5\textwidth]{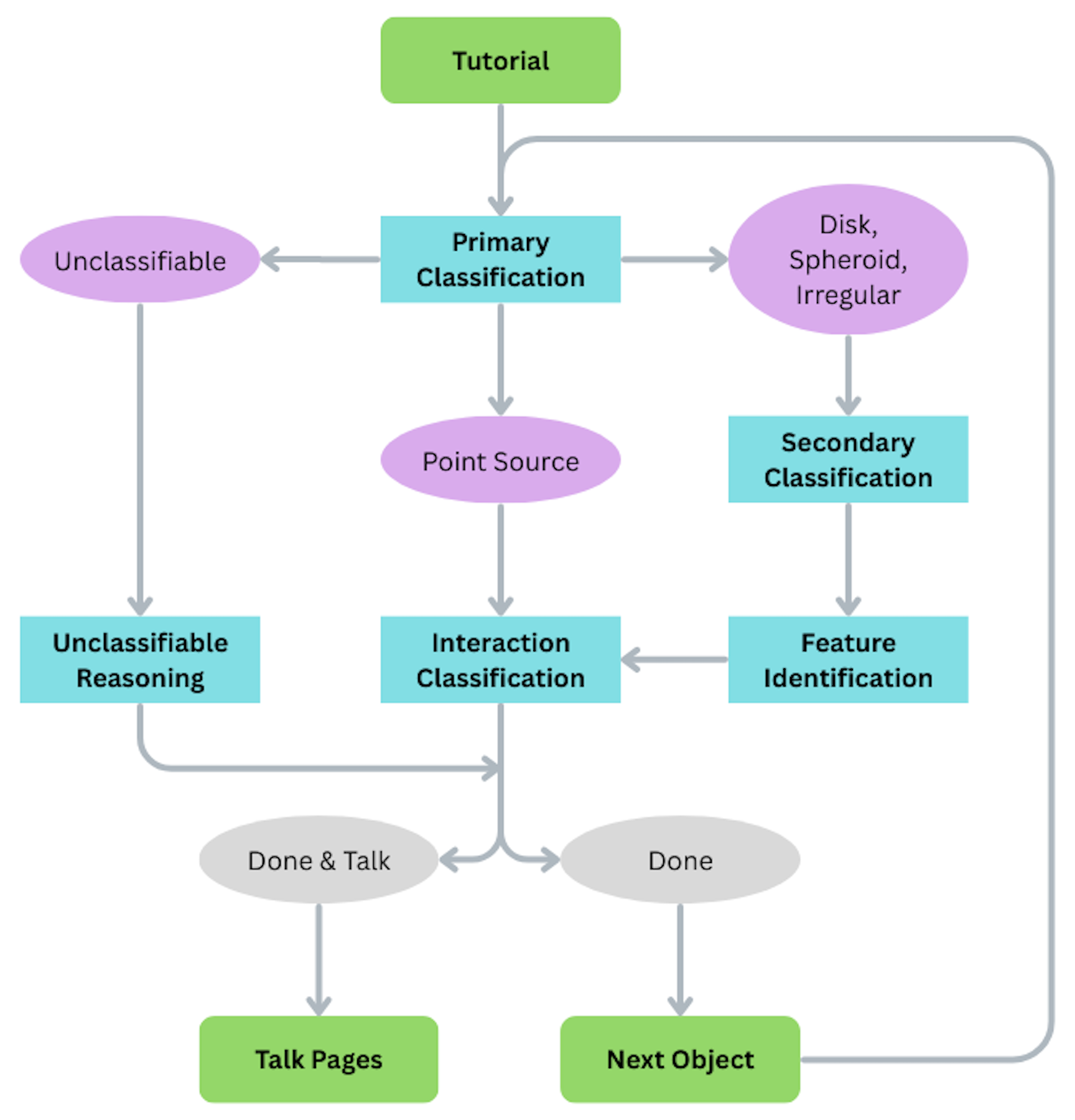}
    \caption{A flowchart showing the questions asked of visual classifiers. Blue rectangles denote Tasks within the workflow. Purple and gray ovals denote responses within our workflow and the Zooniverse user interface, respectively, which result in traversing different branches of the decision tree. Green rounded rectangles denote start/end points.
\label{fig:classification_flowchart}}
\end{figure}

We enlisted the help of citizen science volunteers on Zooniverse\footnote{https://www.zooniverse.org} in order to visually classify our dataset. Zooniverse is a platform which allows the general public to categorize large amounts of data which would be impractical or prohibitively time consuming for individuals or small teams.

\subsubsection{Workflow}\label{subsubsec:workflow}

Figure \ref{fig:classification_flowchart} shows a flowchart describing the classification scheme (or `workflow'), which is composed of five tasks. Our classification scheme is a simplified version of the internal classifications used in \cite{jeyhan2015}. The tasks asked of users are:

\begin{itemize}
    \item \textbf{Task 1: Primary Classification.} Users are shown the text ``What category of galaxy does the object in this image fall under? If the sample could fit multiple categories, pick the primary category.'' Possible responses are: Disk, Spheroid, Irregular, Point Source, or Unclassifiable. Users may only select a single option.
    \item \textbf{Task 2: Secondary Classification.} Users are shown the text ``If the sample could fit multiple categories, pick the secondary category. Otherwise, pick the `N/A' option.'' Possible responses are: Disk, Spheroid, Irregular, or N/A. Users may only select a single option and are not presented with the response that they selected for Task 1.
    \item \textbf{Task 3: Feature Identification.} Users are shown the text ``What features or structures can you identify in the galaxy? Check all that apply. If none apply, choose `N/A'.''. Possible responses are: Spiral, Bar, Ring, Clumps, Bulge, or N/A. Users may select one or more options.
    \item \textbf{Task 4: Interaction Classification.} Users are shown the text ``Which option best describes the interactions that this galaxy is undergoing?'' Possible responses are: Merger, Interacting Companion(s), Non-interacting Companion(s), or None. Users may only select a single option. 
    \item \textbf{Task 5: Unclassifiable Reasoning.} Users are shown the text ``Please select why this image is not classifiable as a galaxy. If you select `Other', please tell us why in the talk pages!'' Possible responses are: Image is only noise / image is empty; The central object is washed out or not present; Image artifact; This is a small piece of a larger galaxy / image is too zoomed in; This is a star, not a galaxy; and Other. Users may only select a single option.
\end{itemize}

Users who choose a resolved morphology (Disk, Spheroid, or Irregular) in Task 1 are shown Tasks 2, 3, and 4. Users who select Point Source are taken directly to Task 4, since an unresolved object cannot reasonably be expected to have secondary structure or visible surface features, but could conceivably be interacting with neighbor. Users who select Unclassifiable are diverted directly to Task 5. 

After users complete the set of Tasks for a single object, they may either move on to a new, randomized object, or they can discuss that object on the talk pages. Our talk pages were divided into two forums: Notes, where users could post general comments or questions on objects that they found interesting; and Science Questions and Discussion, where users could ask more general astronomy questions or ask for guidance in classification from our team. Recurring questions in the talk pages provided insight into ways to refine the resources available to users as the project developed.  

\subsubsection{Tutorial}\label{subsubsec:tutorial}

\begin{figure}[ht]
\includegraphics[width=0.5\textwidth]{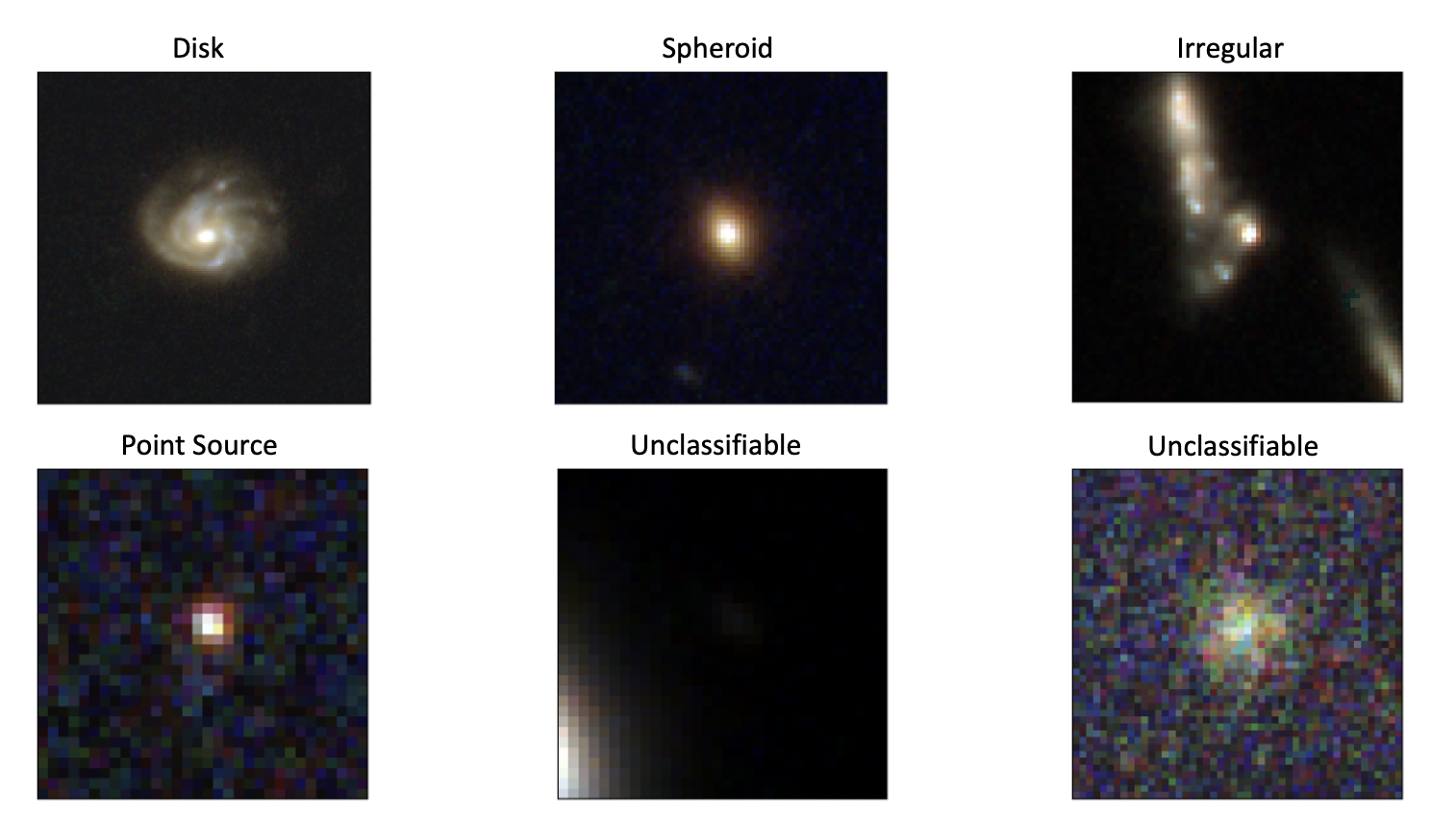}
    \caption{The image shown on the second slide of the tutorial, which explains the main morphology classifications.
\label{fig:tutorial1_mainmorph}}
\end{figure}

\begin{figure}[ht]
\includegraphics[width=0.5\textwidth]{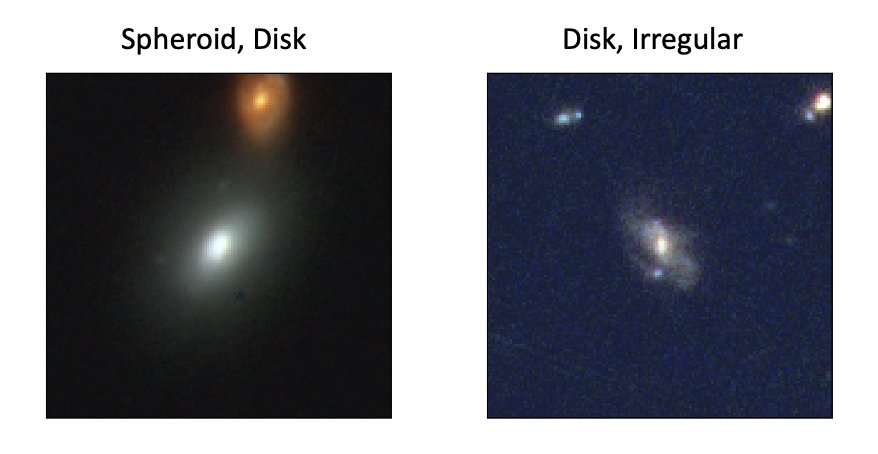}
    \caption{The image shown on the fourth slide of the tutorial, which explains secondary morphology classification.
\label{fig:tutorial2_secondarymorph}}
\end{figure}

\begin{figure}[ht]
\includegraphics[width=0.5\textwidth]{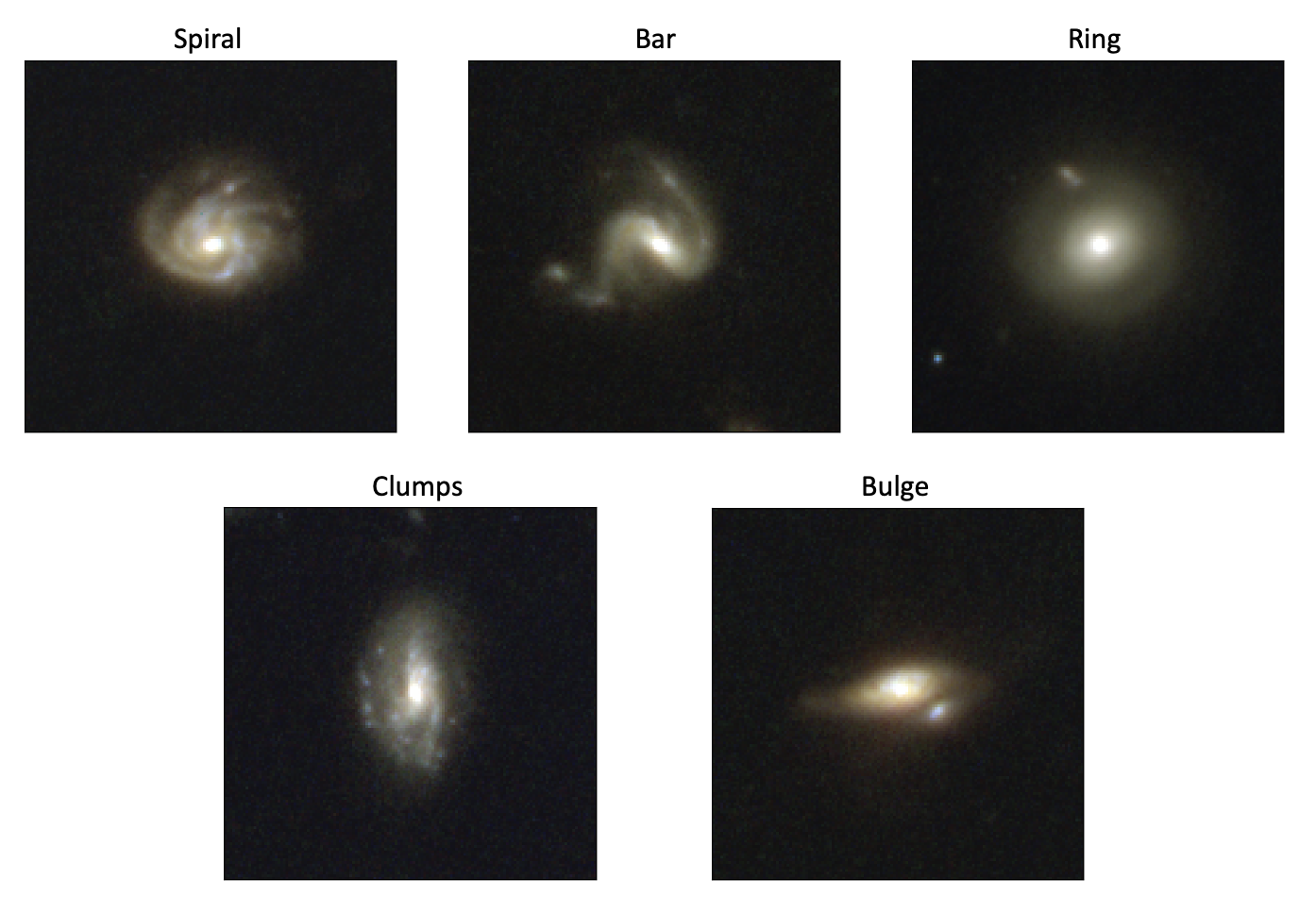}
    \caption{The image shown on the fifth slide of the tutorial, which explains the options for features.
\label{fig:tutorial3_feat}}
\end{figure}

\begin{figure}[ht]
\includegraphics[width=0.5\textwidth]{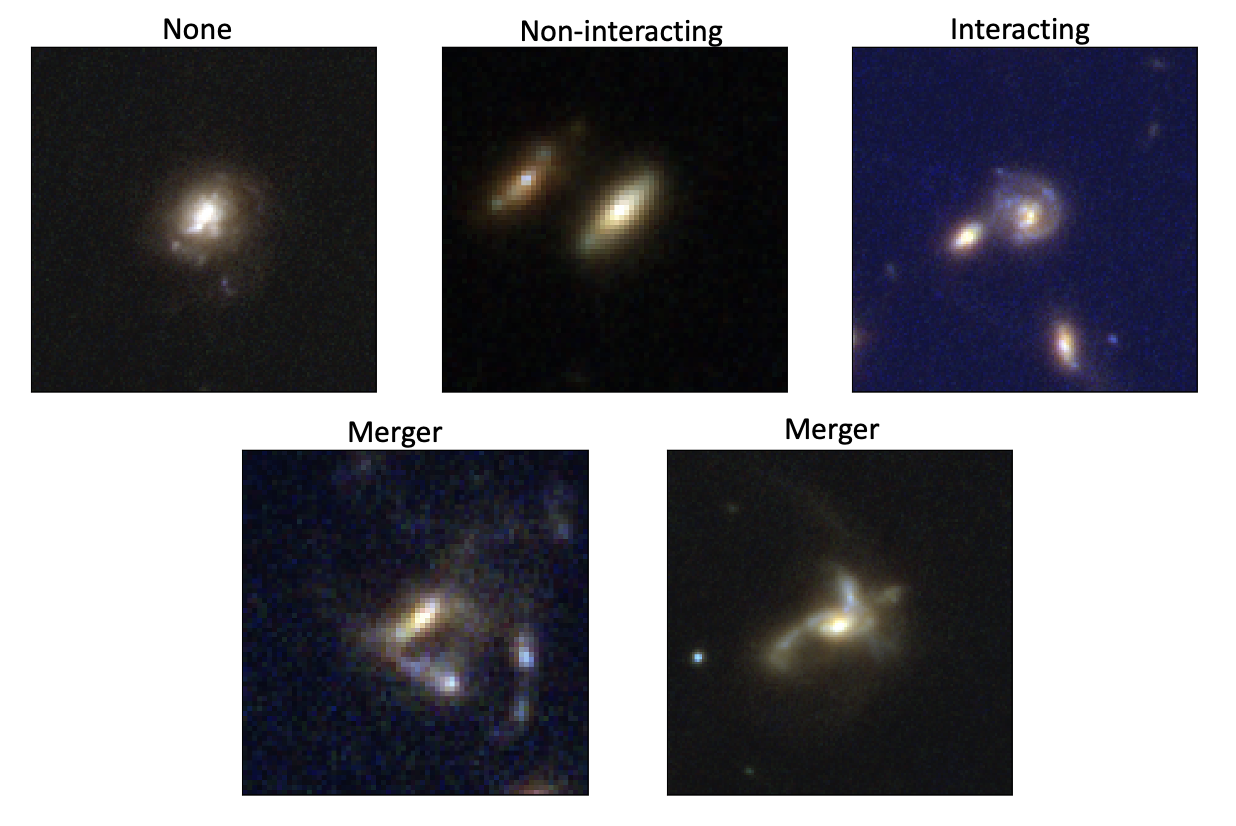}
    \caption{The image shown on the seventh slide of the tutorial, which explains interaction classification.
\label{fig:tutorial4_int}}
\end{figure}

Before users are shown their first object, they are shown a brief tutorial with eight slides. They can also access the tutorial at any point while classifying. The first slide displays the object in Figure \ref{fig:image_example}, explains that it is three images of the same object, and that their goal is to classify the object in the center of the image. The second slide displays the image shown in Figure \ref{fig:tutorial1_mainmorph} and provides definitions of the five primary classifications of Task 1. The third slide provides additional clarification on reasons why a user may choose Unclassifiable for Task 1. The fourth slide shows examples of galaxies which fit multiple categories and explains how to handle secondary morphologies (Figure \ref{fig:tutorial2_secondarymorph}). The fifth slide shows examples of and provides definitions for the features that users are asked to identify in Task 3 (Figure \ref{fig:tutorial3_feat}). The sixth slide prepares users for the possibility that the objects they'll see may not match the pristine examples shown in the tutorial and asks that they do their best to choose a category. The seventh slide shows examples of and provides definitions for the various interaction classes in Task 4 (Figure \ref{fig:tutorial4_int}). The eighth and final slide simply wishes the users luck, thanks them for their help, and reminds them to refer to the talk pages if they find anything of interest. 

In addition to the tutorial, users have access to additional examples through a Field Guide, which is available at all times as a tab on the side of the classification page. All example images are objects from the PRIMER UDS catalog which may or may not have MIRI matches and are created using all available NIRCam bands.

\subsubsection{Beta Test}\label{subsubsec:betatest}

Before the project was publicly launched, we performed a Beta Test with limited public visibility. The Beta Test set consisted of 300 objects from the UDS field, each of which had all available NIRCam bands included in the color composite but was otherwise identical to the images described in \S \ref{subsec:images}. The workflow was the same as described in \S \ref{subsubsec:workflow} but without Task 5. Each object received 30 classifications, and this data was used to test and refine our classification pipeline. 

We also collected user feedback through a survey which was posted in the announcement banner. This survey asked users if they felt the project goals were clear, if the tasks made sense, if sufficient training and reference resources were available, as well as many other opinions on the various aspects of the project. This was used to refine the workflow, reference resources, and information available on the project page before a public launch. 

\subsubsection{Data Uploads}\label{subsubsec:seasons}

The Zooniverse team recommends uploading subjects in batches of a few thousand at a time, so that volunteers can see appreciable progress before moving on to a new data set. Our images were uploaded to Zooniverse in four sets, referred to subsequently as `Seasons'. Season 1 included 3473 objects in the then-incomplete mosaic of the PRIMER UDS field, where objects had a SNR $>$ 10 in all NIRCam bands. Season 2 consisted of 4049 objects in the PRIMER COSMOS mosaic with SNR $>$ 10 in all NIRCam bands, similar to Season 1. Season 3 relaxed the coverage standards to a minimum of two optical-matching NIRCam bands with SNR $>$ 10 in the COSMOS field, which yielded 3436 additional objects. Season 4 used the completed mosaic of the UDS field and the relaxed detection threshold to upload an additional 3433 objects. Each object was retired after 20 classifications, though a small number received more. 

Task 5 was added after Season 1 was processed, and we found that users were sometimes selecting Unclassifiable for objects which were poorly resolved or had low surface brightness and therefore only \textit{difficult} to classify, not impossible. By giving users a list of reasons that they should be marking objects as Unclassifiable (i.e. bad data), we attempted to prevent this particular source of `noise' on the classification data for subsequent seasons.

Due to improvements in the reduction and catalog production as methods were refined, the UDS mosaic used to generate the Beta Test/Season 1 data differed from that used for Season 4. We reconciled the two catalogs and found that 34 objects from the Beta Test were not included in the updated version; these objects are largely unclassifiable, spurious detections. As such, the total number of objects that were uploaded is 14,691, while our final classification catalog only contains 14,657 objects. 

\subsection{Classification Pipeline}\label{subsubsec:pipeline}

\begin{figure}[h!]
\includegraphics[width=0.5\textwidth]{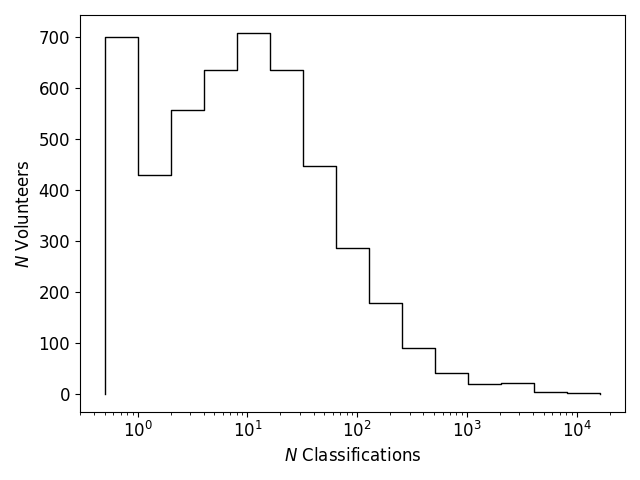}
    \caption{The distribution of number of classifications by identifiable volunteers.  Each bin is twice the width of the previous bin. Many users submitted only a single classification, while most submitted a few dozen. Some, however, submitted hundreds or even thousands. 
\label{fig:classcounts}}
\end{figure}

Cosmic Collisions accumulated 304,524 classifications for 14,691 objects. We had a minimum of 4742 unique classifiers (users who created an account), each of whom contributed, on average, 59.58 classifications, but with significant variance. 21,978 (7.2\% of the total) classifications were submitted by an unknown number of anonymous users. Figure \ref{fig:classcounts} shows the distribution of classification submissions by our identifiable volunteers. 700 volunteers submitted only a single classification. The typical user submitted between 9 and 16 classifications. There were 44 dedicated volunteers who submitted over 1000 classifications apiece, accounting for $37.6\%$ of our data.

We determined final classifications through an iterative scheme that takes into account the fractional breakdown of votes, the number of classifications each user submitted, and the agreement among classifiers. Stage 1 calculates a set of raw fractions for each object. Stage 2 estimates reliability weightings for each user. Stage 3 uses these weights to calculate weighted versions of the fractions created in Stage 1. Stage 4 iterates on Stages 2 and 3 until convergence. Stage 5 uses the weighted fractions to apply a final set of classifications to each object.

\subsubsection{Stage 1: Raw Fractions}

We check the simple fraction of classifiers that selected an option $p$ according to Equation \ref{eq:fraction}:
\begin{equation}
    p = \frac1n\sum_{i=1}^{n} s_iw_i
\label{eq:fraction}
\end{equation}
where $n$ is the total number of classifiers for that object, $s_i$ is 1 if the classifier selected the option and 0 if they did not, and $w_i$ is the weight given to that classification. Tasks 1 and 2 are considered simultaneously. If no secondary classification was selected in Task 2, $w_i=1$ for the response in Task 1. If a secondary classification was selected, $w_i=0.67$ for the primary classification and $w_i=0.33$ for the secondary classification. All other Tasks are considered independently, and for all other options and Tasks, $w_i=1$.

\subsubsection{Stage 2: Reliability}\label{subsubsec:p2}

As noted in \cite{galaxyzoo}, a simple fraction does not differentiate between users who classify more thoughtfully than others, nor does it differentiate between users of varying ability or background knowledge. We therefore weight the classification of each user by how often their votes agree with the majority. We construct four agreement fractions for each user: one for Tasks 1 and 2 and one for each of Tasks 3 through 5. This approach has the advantage of distinguishing between users who may excel at one Task but not another. For each user and Task (or set of Tasks), we record the fraction of responses which agree with the majority according to Equation \ref{eq:relfrac},

\begin{equation}
    f = \frac{a}{N_e}
\label{eq:relfrac}
\end{equation}

\noindent where $f$ is the agreement fraction for a given user and task, $N_e$ is the number of times the user encountered that Task, and $a$ is the number of times they voted with the majority on that Task. For Tasks 1 and 2, a response is only considered full agreement, increasing $a$ by 1, if the user voted with the majority in Task 1. If they vote with the majority in Task 2 (i.e.\ voting that the consensus primary morphology is secondary), then we instead increase $a$ by $\frac12$.  For Task 3, if the user's answer includes the most-voted feature, including None, this is considered agreement. If a user never encounters a Task, then $N_e = 0$ and no weight is assigned. This generally only happened for Task 5. 

 This simple agreement fraction can create the false impression that a user who submitted one `correct' classification is more reliable than a user who submitted many. Users who submit a greater number of classifications can reasonably be assumed to be well-practiced, giving higher quality responses. This must be considered as well. To get a final reliability weighting, we multiply each agreement fraction by a classification count weight $C_w$, calculated using Equation \ref{eq:logClassCount}:

\begin{equation}
    C_w = A\ \text{log}_{10}(N)^\alpha + c
\label{eq:logClassCount}
\end{equation}

\begin{equation}
    r = C_wf
\label{eq:finalrel}
\end{equation}

\noindent where $r$ is the reliability weighting, and $N$ is the number of classifications submitted by the user. $A$, $c$, and $\alpha$ are arbitrary constants, determined so $C_w$ has a value of 0.5 when a user submits only one classification, a value of 1.0 when a user submits the average number of classifications $\bar N$, and a value of 1.5 for the highest number of classifications submitted by any user $N_{\rm max}$. The functional form of Equation \ref{eq:logClassCount} is constructed such that it spans the wide range of classifications and bends to hit these three benchmarks exactly. With $\bar N= 59.58$ and $N_{\rm max}=10230$, $A = 0.304$, $c=0.5$, and $\alpha=0.873$.

For users who submitted classifications anonymously, individual reliability scores cannot be distinguished, but reliability weightings are still necessary. For these classifications, we calculate an agreement fraction as described above on all anonymous classifications, which we interpret as the average reliability of unregistered users. Because the number of classifications attributable to individuals is unknown, we assume that each classification is submitted by a unique user and assign $C_w=0.5$ accordingly.

\begin{figure}[h!]
\includegraphics[width=0.5\textwidth]{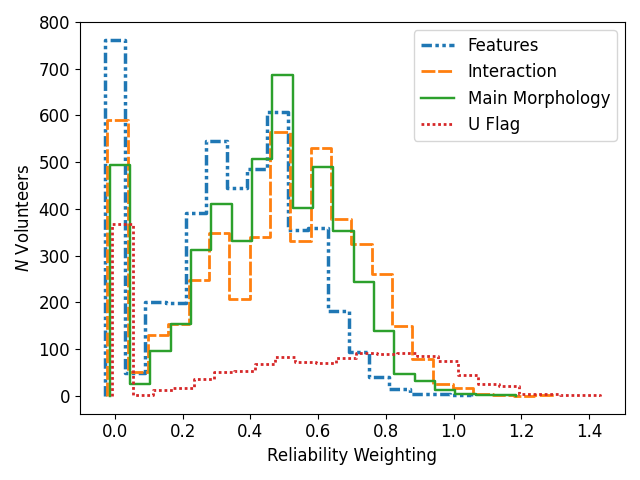}
    \caption{The reliability weights of identifiable volunteers. Main morphology reliability (Tasks 1 and 2) is shown as the green solid line, interaction reliability (Task 4) is shown as the orange dashed line, feature reliability (Task 3) is shown as the blue dot dashed line, and unclassifiable flag reliability (Task 5; abbreviated as U flag) is shown as the red dotted line. Note that these are the post-iteration reliability weightings produced in Stage 4.
\label{fig:reliability}}
\end{figure}

Figure \ref{fig:reliability} shows the distribution of reliability weightings. In general, users agree on an object's classification $\sim$$60\%$ of the time. The average reliability is around 50\%, as this takes into account the number of classifications as well. There are 1188 users who had their votes down-weighted to zero in at least one category; this results in less than 1\% of total classifications being disregarded in some way.

\subsubsection{Stage 3: Weighted Fractions}\label{subsubsec:p3}

We calculate a weighted classification fraction using the reliability weightings according to Equation \ref{eq:weightedFraction}:
\begin{equation}
    p_w = \frac{\sum_{i=1}^{n} s_i w_i r_i}{\sum_{i=1}^{n} r_i}
\label{eq:weightedFraction}
\end{equation}

\noindent where $s_i$ and $w_i$ are defined the same as in Equation \ref{eq:fraction}, and $r_i$ is the reliability of each user for the category being calculated. The sum of vote fractions for Tasks 1-2 and Task 4 will always be 1.0; if an object is flagged as unclassifiable in a Season of data which includes Task 5, these votes will sum to 1.0 as well. Because Task 3 allows users to select any number of options, the sum of Task 3 responses will not always sum to 1.0.

\subsubsection{Stage 4: Iteration}\label{subsubsec:p4}

The calculation of the weighted classification fractions can change which option received the majority of weighted votes. This, in turn, means that the reliabilities generated in the previous step are no longer accurate. After the weighted fractions are calculated, we use these to recalculate reliability weights, which are, in turn, used to calculate new weighted classifications. This process is iterated until convergence. The end result is a table of users with reliability weights for each Task and a table of raw and weighted fractional classifications for each object. 

\subsubsection{Stage 5: Consensus and Subsets}\label{subsubsec:p5}

We use the fractional weighted classification to determine a single descriptor for each object and Task as follows: for Tasks 1-2, if the highest rated option $p_{w,1}$ received a majority ($p_{w,1} > 0.5$) of votes, we report that option. If no option received a majority, we check the second-highest option, $p_{w,2}$, and determine if $p_{w,1} + p_{w,2}> 0.667$; this threshold is chosen as the lowest value which can satisfy $p_{w,2} > \frac{1-p_{w,1}}2$. If this condition is met, we report $p_{w,1}$ as the primary morphology and $p_{w,2}$ as secondary. Objects which meet neither condition are marked having inconclusive morphology. Objects which mix resolved (disk, spheroid, irregular) and unresolved (point source, unclassifiable) options are flagged as unreliable (the q\_flag in Table \ref{tab:visfracs}). 

For Task 3, we report any feature which a majority of classifiers voted as present. If 68\% of classifiers selected `N/A' for features, all other features are overridden. In practice, this can only happen if multiple users vote that features are both present and not present. If no feature has $p_{w}>0.5$, including `N/A', the object is marked as having inconclusive features. 

For Task 4, the `Merger' or `Interacting Companion(s)' choices both indicate that the object is undergoing or has recently undergone some form of disturbance. We find that these two options often dilute one another, leading to difficulty drawing a conclusion. We therefore reduce this Task to two possible results, `Interacting' or `Non-interacting'. We conclude the former if the sum of `Merger' and `Interacting Companion(s)' votes is greater than 0.667 and the latter if the sum of `Non-interacting Companion(s)' and `None' votes is greater than 0.667. If neither threshold is cleared, the object is marked as inconclusive.

For Task 5, if the primary or secondary morphology for Tasks 1-2 is selected as `Unclassifiable' (and the object is in a Season which includes Task 5) then the object is marked with whichever Task 5 option received the most votes. Since inconclusivity as a descriptor is primarily useful for pruning the data set, doing so on an object already removed for being Unclassifiable is unnecessary. As such, it is more informative to record why this option was selected, regardless of whether the classifiers are in consensus. If the object is in Season 1 or the Beta Test, or if Unclassifiable is not among the chosen descriptors for Tasks 1 and 2, the conclusion for Task 5 is marked as `n/a'. 

Broadly, the data product of interest is the fractional classification data, and reducing this to a single identifier in each category is a significant loss of information, although it may be more useful when considering population trends. The process of identifying inconclusive or questionable results is useful for delineating subsets by quality. Some useful quality subsets are defined in Table \ref{tab:subsets}, and the calculations described above are shown mathematically in Table \ref{tab:pops}. We include these descriptors in Table \ref{tab:visfracs}, but we encourage users to define subsets, confidence thresholds, and conclusions to suit their own needs.

\section{Results and Discussion}\label{sec:results}

\subsection{Data Presentation and Redshift Evolution}\label{subsec:visdata}

\setlength{\LTcapwidth}{\textwidth}

\begin{longtable*}[c]{|c c|c c c c c c c|}
\caption{
    Raw and weighted classification fractions for the entire Cosmic Collisions Sample. Full table available for download at [URL GOES HERE]. The printed table is a transposed version (the columns appear as rows here for easier formatting within this paper) with examples and numbered columns to illustrate format. The complete version contains all 14,657 objects. Columns 1 and 2 give the object's numerical ID in the internal PRIMER catalog and the field. Columns 3 and 4 give its coordinates, and column 5 gives its redshift. Column 6 gives the number of classifications submitted for each object. Columns 7-11, 12-17, 18-21, and 22-27 give the raw fractional vote breakdown for each possible answer of the Morphology, Feature, Interaction, and Unclassifiable reasoning questions, respectively. Columns 28-31 give the sum of user weights which went into classifying the object for each question. Columns 32-52 are correspond to columns 7-27, recalculated using weighted votes and are distinguished using with the `\_w' suffix. Columns 53, 54, 55, and 56 are the conclusions for each of the Morphology, Feature, Interaction, and Unclassifiable reasoning questions, respectively. Column 57 is a data quality flag which indicates when votes yielded a mix of resolved and unresolved morphologies. Lastly, column 58 is a numerical subset indicator which is described in Table \ref{tab:subsets}. Specific definitions of column names are provided in the header of the published table.)
}\label{tab:visfracs} \\

\hline 
\multicolumn{1}{|c}{} &
\multicolumn{1}{c|}{\textbf{Column Name}} & \multicolumn{7}{c|}{\textbf{Subject}} \\ 
\hline 
\endfirsthead

\multicolumn{9}{c}%
{{\bfseries \tablename\ \thetable{} -- continued from previous page}} \\
\hline
\endhead
\hline \multicolumn{9}{|r|}{{Continued on next page}} \\ \hline
\endfoot
\hline \hline
\endlastfoot
\centering
1 & id & 17715 & 17755 & 17746 & 4311 & 4568 & 5558 &  ... \\ 
2 & field & UDS & UDS & UDS & COSMOS & COSMOS & COSMOS &  ... \\ 
\hline
3 & ra & 34.290698 & 34.403252 & 34.464085 & 150.143706 & 150.135748 & 150.136763 &  ... \\ 
4 & dec & -5.294295 & -5.294458 & -5.29439 & 2.153892 & 2.154537 & 2.156497 &  ... \\ 
5 & z & 5.485072 & 2.016344 & 2.745998 & 1.896218 & 3.912955 & 1.670523 &  ... \\ 
6 & nclass & 21 & 20 & 20 & 30 & 25 & 31 &  ... \\ 
7 & m\_disk & 0.047619 & 0.05 & 0.3325 & 0.555333 & 0.1064 & 0.214839 &  ... \\ 
8 & m\_sphere & 0.079524 & 0.0 & 0.351 & 0.067 & 0.4004 & 0.097097 &  ... \\ 
9 & m\_irr & 0.063333 & 0.85 & 0.0165 & 0.311 & 0.0132 & 0.204194 &  ... \\ 
10 & m\_point & 0.666667 & 0.0 & 0.3 & 0.033333 & 0.44 & 0.322581 &  ... \\ 
11 & m\_unc & 0.142857 & 0.1 & 0.0 & 0.033333 & 0.04 & 0.16129 &  ... \\ 
12 & f\_spiral & 0.0 & 0.0 & 0.2 & 0.233333 & 0.04 & 0.064516 &  ... \\ 
13 & f\_bar & 0.047619 & 0.05 & 0.0 & 0.166667 & 0.0 & 0.032258 &  ... \\ 
14 & f\_ring & 0.0 & 0.05 & 0.25 & 0.133333 & 0.04 & 0.064516 &  ... \\ 
15 & f\_clump & 0.0 & 0.6 & 0.0 & 0.066667 & 0.04 & 0.064516 &  ... \\ 
16 & f\_bulge & 0.047619 & 0.1 & 0.3 & 0.566667 & 0.16 & 0.193548 &  ... \\ 
17 & f\_none & 0.904762 & 0.3 & 0.45 & 0.233333 & 0.72 & 0.709677 &  ... \\ 
18 & i\_merger & 0.0 & 0.25 & 0.05 & 0.0 & 0.0 & 0.064516 &  ... \\ 
19 & i\_intcom & 0.047619 & 0.3 & 0.05 & 0.133333 & 0.08 & 0.096774 &  ... \\ 
20 & i\_nintcom & 0.0 & 0.3 & 0.1 & 0.066667 & 0.0 & 0.0 &  ... \\ 
21 & i\_none & 0.809524 & 0.05 & 0.8 & 0.766667 & 0.88 & 0.677419 &  ... \\ 
22 & u\_noise & 0.0 & 0.0 & 0.0 & 0.0 & 0.0 & 0.4 &  ... \\ 
23 & u\_noObj & 0.0 & 0.0 & 0.0 & 0.0 & 0.0 & 0.4 &  ... \\ 
24 & u\_artifact & 0.0 & 0.0 & 0.0 & 0.0 & 0.0 & 0.0 &  ... \\ 
25 & u\_zoom & 0.0 & 0.0 & 0.0 & 1.0 & 1.0 & 0.0 &  ... \\ 
26 & u\_star & 0.0 & 0.0 & 0.0 & 0.0 & 0.0 & 0.0 &  ... \\ 
27 & u\_other & 0.0 & 0.0 & 0.0 & 0.0 & 0.0 & 0.2 &  ... \\ 
28 & weight\_morph & 13.572443 & 14.423681 & 12.233113 & 17.963669 & 15.613725 & 20.108084 &  ... \\ 
29 & weight\_feat & 10.641644 & 11.732252 & 9.737318 & 14.916608 & 14.016138 & 17.476966 &  ... \\ 
30 & weight\_int & 15.721269 & 14.706444 & 13.774568 & 21.16553 & 20.270682 & 25.293848 &  ... \\ 
31 & weight\_unc & 5.550539 & 7.430463 & 3.053323 & 16.633441 & 16.108934 & 19.181419 &  ... \\ 
32 & m\_disk\_w & 0.012969 & 0.022622 & 0.321586 & 0.617872 & 0.105687 & 0.262899 &  ... \\ 
33 & m\_sphere\_w & 0.105894 & 0.0 & 0.39548 & 0.070746 & 0.52257 & 0.108735 &  ... \\ 
34 & m\_irr\_w & 0.042388 & 0.905275 & 0.019619 & 0.264261 & 0.01859 & 0.212301 &  ... \\ 
35 & m\_point\_w & 0.718632 & 0.0 & 0.263315 & 0.028552 & 0.331788 & 0.251991 &  ... \\ 
36 & m\_unc\_w & 0.120116 & 0.072104 & 0.0 & 0.018569 & 0.021364 & 0.164074 &  ... \\ 
37 & f\_spiral\_w & 0.0 & 0.0 & 0.207316 & 0.137393 & 0.012266 & 0.063132 &  ... \\ 
38 & f\_bar\_w & 0.018397 & 0.016687 & 0.0 & 0.121841 & 0.0 & 0.009837 &  ... \\ 
39 & f\_ring\_w & 0.0 & 0.016687 & 0.263299 & 0.09743 & 0.018121 & 0.028785 &  ... \\ 
40 & f\_clump\_w & 0.0 & 0.650356 & 0.0 & 0.072669 & 0.037211 & 0.083544 &  ... \\ 
41 & f\_bulge\_w & 0.05368 & 0.129275 & 0.271227 & 0.551467 & 0.172957 & 0.185432 &  ... \\ 
42 & f\_none\_w & 0.927923 & 0.267429 & 0.445623 & 0.279051 & 0.759445 & 0.743588 &  ... \\ 
43 & i\_merger\_w & 0.0 & 0.225013 & 0.015725 & 0.0 & 0.0 & 0.033406 &  ... \\ 
44 & i\_intcom\_w & 0.01186 & 0.36451 & 0.020343 & 0.069629 & 0.045884 & 0.05407 &  ... \\ 
45 & i\_nintcom\_w & 0.0 & 0.274314 & 0.110195 & 0.022546 & 0.0 & 0.0 &  ... \\ 
46 & i\_none\_w & 0.888342 & 0.041707 & 0.853738 & 0.87892 & 0.923934 & 0.78466 &  ... \\ 
47 & u\_noise\_w & 0.0 & 0.0 & 0.0 & 0.0 & 0.0 & 0.329621 &  ... \\ 
48 & u\_noObj\_w & 0.0 & 0.0 & 0.0 & 0.0 & 0.0 & 0.498289 &  ... \\ 
49 & u\_artifact\_w & 0.0 & 0.0 & 0.0 & 0.0 & 0.0 & 0.0 &  ... \\ 
50 & u\_zoom\_w & 0.0 & 0.0 & 0.0 & 1.0 & 1.0 & 0.0 &  ... \\ 
51 & u\_star\_w & 0.0 & 0.0 & 0.0 & 0.0 & 0.0 & 0.0 &  ... \\ 
52 & u\_other\_w & 0.0 & 0.0 & 0.0 & 0.0 & 0.0 & 0.172091 &  ... \\ 
53 & c\_morph & point source & irregular & spheroid+disk & disk & spheroid & inconclusive &  ... \\ 
54 & c\_feat & n/a & clump & inconclusive & bulge & none & none &  ... \\ 
55 & c\_int & non-interacting & inconclusive & non-interacting & non-interacting & non-interacting & non-interacting &  ... \\ 
56 & u\_flag & n/a & n/a & n/a & n/a & n/a & n/a &  ... \\ 
57 & q\_flag & 0 & 0 & 0 & 0 & 0 & 0 &  ... \\ 
58 & subset\_flag & 1 & 24 & 23 & 1 & 1 & 34 &  ... \\ 

\end{longtable*}

\begin{table}
\begin{center}
\caption{Definition of data quality subsets and their associated flags in column 58 of Table \ref{tab:visfracs}, and the number of objects in each subset. Conclusivity, primary, and secondary morphologies in this table are defined in \S \ref{subsubsec:pipeline}. Clean and Superclean usurp all other quality flags. Flags 2-4 are appended to one another; for example, an object with a conclusive morphology classification and feature classification, but not a conclusive interaction classification, would be denoted as `subset\_flag = 24', while one with only a conclusive interaction would be marked as `subset\_flag = 3'. Primary and Secondary morphologies are abbreviated as P. or S. \label{tab:subsets}}

\begin{tabular}{|l|l|l|l|}
\hline
Subset & Definition & $N$ & Flag \\ 
\hline \hline 
Superclean & Conclusive in all categories & 1147 & 0 \\
           & 0 votes for Unclassifiable & & \\
           & 0 votes for Point Source & & \\
\hline
Clean & Conclusive in all categories  & 4880 & 1 \\
      & Not P. or S. Unclassifiable & & \\ 
      & Not S. Point Source & &  \\ 
      & Not P. Point Source + S. Any & &  \\ 
\hline
Good Morph & Conclusive Morphology & 5174 & 2 \\
 & Not P. or S. Unclassifiable & & \\
 & Not S. Point Source & &  \\
 & Not P. Point Source + S. Any & &  \\ 
 \hline
Good Int & Conclusive Interactions & 4545 & 3 \\
 & Not P. Unclassifiable & & \\
\hline
Good Feat & Conclusive Features & 3526 & 4 \\
 & Not P. Unclassifiable & & \\
 & Not P. or S. Point Source & & \\
\hline
Bad Data & Not in another subset & 646 & 50 \\
\hline
\end{tabular}
\end{center}
\end{table}

\begin{figure}[h!]
\includegraphics[width=0.5\textwidth]{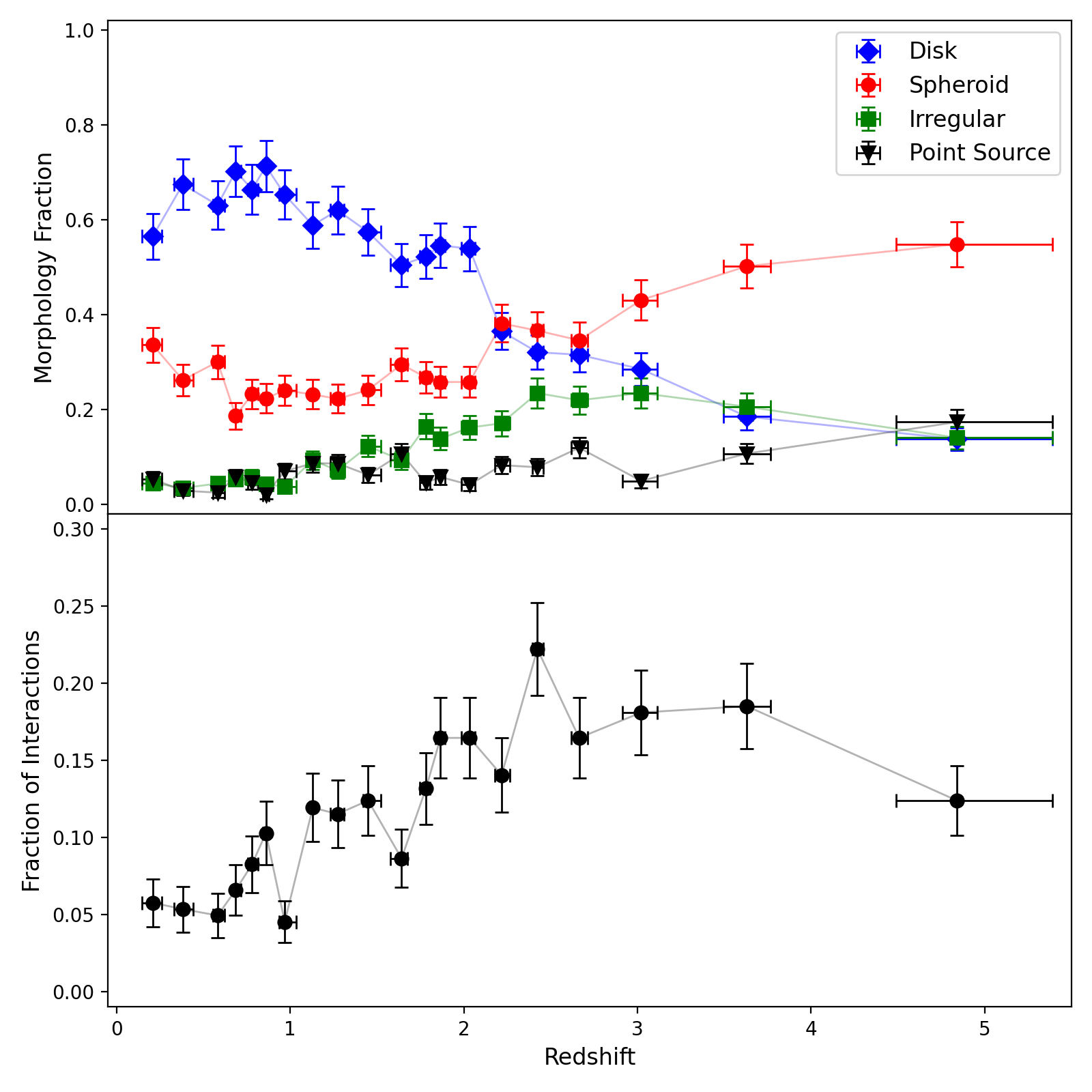}
    \caption{The main morphology and interaction fractions of galaxies as a function of redshift for objects in the Clean and Superclean subsample. We plot 20 redshift bins containing equal numbers of objects. Vertical error bars are Poisson errors on population counts. The plotted point is the median redshift in the bin, and the horizontal error bars show the upper and lower quartile of the bin.}
\label{fig:morphofz}
\end{figure} 

Table \ref{tab:visfracs} shows the raw and weighted fractional classification of the Cosmic Collisions project. The full table can be accessed at [URL]. Table \ref{tab:subsets} defines various data quality subsets which are flagged in the `subset\_flag' column of Table \ref{tab:visfracs}, included for ease of use. These are arranged such that data quality can be selected with a simple value cut. Cleanly classified objects can be selected with `subset\_flag $<$ 2'; data with inconclusive marked in some categories will have a subset\_flag value between 2 and 45; and bad data can be excluded with `subset\_flag $<$ 50'.

\begin{table}
\begin{center}
\caption{Breakdown of various populations within the Clean and Superclean subsamples. Spheroid is abbreviated to sph, Irregular is abbreviated to irr,  Point source is abbreviated to point, Interaction Companion(s) is abbreviated to int, and Non-interacting Companion(s) is abbreviated to nint. None refers to an object which is alone in the frame (i.e., no companions of any kind). $f$ describes an arbitrary weighted morphology fraction, with subscripts denoting specific fractions.\label{tab:pops}}
\begin{tabular}{|l|l|l|}
\hline
Population & Definition & $N$  \\
\hline \hline
Disk & $f_\text{disk} \ge 0.5$ & 2667 \\
Spheroid & $f_\text{sph} \ge 0.5$ & 1329 \\ 
Irregular & $f_\text{irr} \ge 0.5$ & 499 \\ 
Point Source & $f_\text{point} \ge 0.5$ &  472 \\

Disk+Sph & $(f_{\rm max} = f_{\rm disk} \vee f_{\rm sph})$ & 618 \\
& \& $(f_{\rm disk} +f_{\rm sph}>\frac23)$ & \\

Disk+Irr & $(f_{\rm max} = f_{\rm disk} \vee f_{\rm irr})$ & 299 \\
& \& $(f_{\rm disk} +f_{\rm irr}>\frac23)$ & \\

Sph+Irr & $(f_{\rm max} = f_{\rm sph} \vee f_{\rm irr})$ & 139 \\
& \& $(f_{\rm sph} +f_{\rm irr}>\frac23)$ & \\

\hline
Interacting & $f_\text{merger}+f_\text{int} > \frac23$ & 675 \\
Non-interacting & $f_\text{nint}+f_\text{none} >\frac23$ & 5348 \\
\hline 
Featured & $f_\text{any feature} \ge 0.5$ \& $f_\text{point}<0.5$& 2796 \\
\hline
\end{tabular}
\end{center}
\end{table}

Table \ref{tab:pops} shows the population breakdown of the Clean and Superclean subsample. Throughout the remainder of this paper, we use only the combined Clean and Superclean subsamples in our analysis, unless otherwise specified.

Figure \ref{fig:morphofz} shows how morphology and interaction fractions evolve with redshift. The majority of our sample are non-interacting disks between $1<z<2$. Above $z\sim 3$, we begin to suffer from small number statistics; above $z\sim5.5$, there are too few galaxies to continue to plot trends with any degree of confidence. 

Our data shows that galaxies in the nearby universe tend to be overwhelmingly disk-like, which is in conflict to the established transition from disk-dominated spiral galaxies into bulge-dominated elliptical galaxies in the local universe. This is likely a selection effect; elliptical galaxies tend to be poor in gas - and therefore dust - leading to an underrepresentation in our mid-IR selected sample.

%It is also important to note that the COSMOS and UDS fields are so-called `blank fields', so dense galactic environments are underrepresented. As such, any assertions about this population should be viewed as limited in scope to mid-IR detectable field galaxies at $z\le 4$. (TODO - put this elsewhere? )

Irregular and interacting galaxies increase in frequency until $z\sim 3$, before falling off again. In a review of the importance of mergers in galaxy evolution, \cite{conselice2008_merger} observes that, among the most massive (log$(M/M_\odot)>10^{10}$) galaxies, the merger rate is as high as $40\pm10\%$ at $z\sim3$. A first look at our data shows that while the peak of interactions is somewhere between $z=2.5-3.5$, it is not nearly as high. However, it is important to note that, due to \textit{JWST}'s sensitivity, the objects detected are, on average, extremely faint in comparison to past surveys \citep{kirkpatrick2023}. This raises the possibility that our disagreement with the maximum of interactions may be due to differences in mass.

\begin{figure*}[t!]
    \includegraphics[width=\textwidth,]{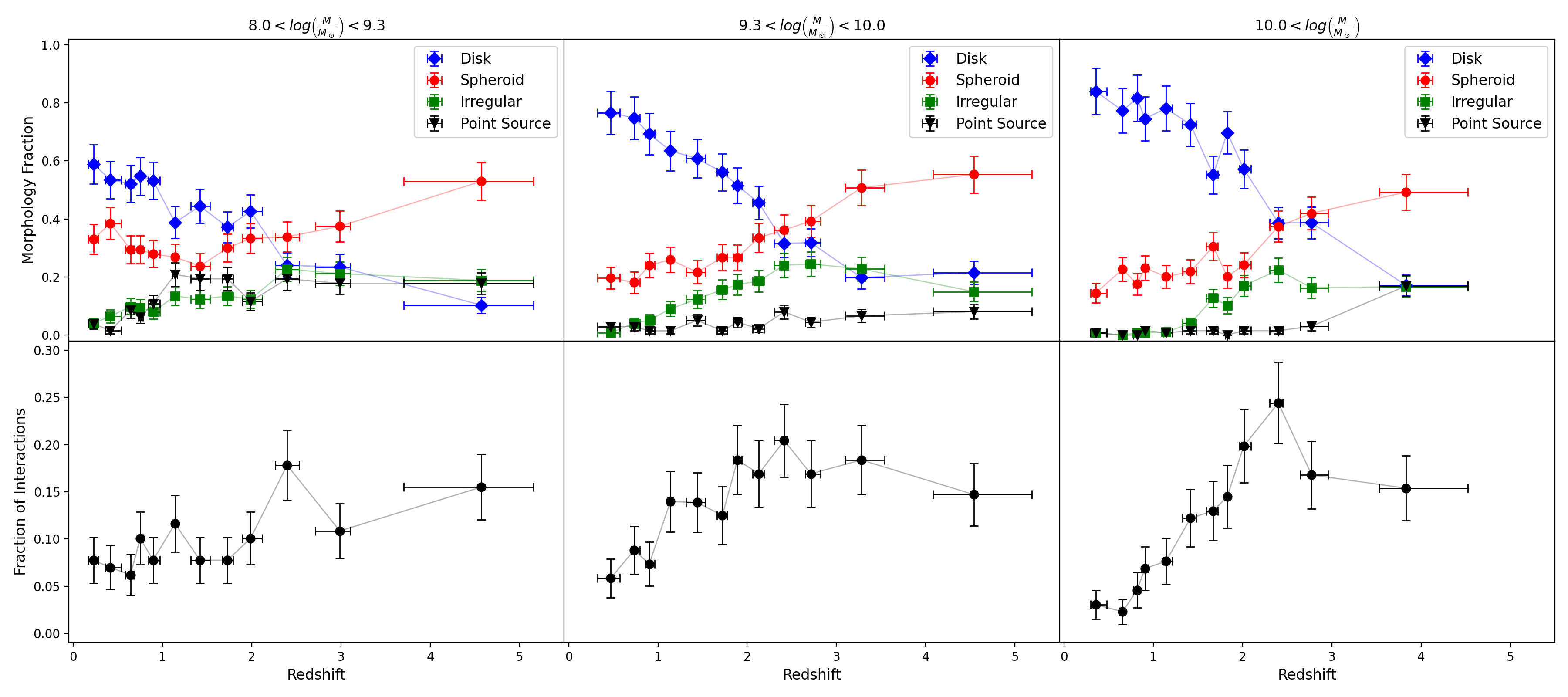}
    \caption{Three panels of Figure \ref{fig:morphofz}, split into lower, middle, and upper 33\% percentile bins by stellar mass, with a lower bound of $M_\odot>10^8$ to prevent skewing at $z\sim0$. The mass range is shown above each column. This shows that the interaction rate peak increases with increasing stellar mass, while the broad trends in morphology are largely unaffected. 
\label{fig:morphofz_massbins}}
\end{figure*}

Figure \ref{fig:morphofz_massbins} shows roughly the same mass bins as used in Figure 14 of \cite{conselice2008_merger}. Even when accounting for stellar mass, the highest rate of interactions in our sample is roughly 25\%; while there is a slight overlap between uncertainties, we must conclude that we observe a lower rate of interactions among massive galaxies. We do find that higher mass galaxies undergo interactions at higher rates. However, our peak of interactions is also constant at $z\sim2.5$, whereas \cite{conselice2008_merger} shows the peak epoch of interactions moving to higher $z$ with increasing mass. The intermediate and low mass bins do largely agree between works. It is important to note the differences in methodology; \cite{conselice2008_merger} identifies mergers using Concentration and Asymmetry statistics after applying a wavelength correction while we identify them directly. Because galaxies become more clumpy and peculiar at high redshift \citep[e.g.][]{abraham1996morphologies, conselice2000_assymetry}, its possible that some systems are being misidentified. However, it is possible our lower peak may simply be an effect of cosmic variance, or systematic effects introduced by our processing pipeline. While this disparity in interaction rate may show that mid-IR selection biases against interaction, we consider this possibility unlikely, as \cite{Kartaltepe_2012} shows that IR-bright galaxies tend to have \textit{higher} interaction rates than other samples. 

It is notable that, in all mass bins, our sample shows that spheroids dominate at high redshift. This is in conflict with evidence from \cite{ceersKartaltepe2023} showing galaxies with disks and irregular features are dominant over spheroids at $3<z<5$ for galaxies with $M_\odot>10^{9}$. These classifications are highly similar to ours (performed in the restframe optical with similar visual classification schemes) and so may be considered directly comparable; therefore discrepancies must be due to intrinsic differences between the populations studied. We see similar disagreement with low redshift results from CEERS \citep{huertas-company2024_jwstmorpho}, though their method uses a CNN and a separate category for Bulge+Disk, so the two works are less easily compared. We explain these differences as the confluence of two effects: first, that citizen science classifiers struggle to differentiate faint point sources from resolved spheroids, due to the lack of identifying PSF features; and second, that our sample is selected using mid-IR matching, and the AGN fraction of mid-IR samples increases with redshift \citep{hamblin2026}. Taken together, this suggests that many high-$z$ `spheroids' are active galaxies with faint but detectable central engines and undetected hosts. This possibility will be investigated further in a future paper. 

% (TODO: further analysis of interest). 

% Tabulate nearest neighbors within redshift uncertainty for each object and see how the disk vs spheroid fraction changes based on proximity to other galaxies --> done but beyond the scope of this work

% See how the pixel sizes of spheroids and point sources compare at high z - does this support the claim that classifiers are confusing point sources and spheroids? --> hard to say but generally, size vs classification fraction for point sources traces that of spheroids. does support.

% see what the breakdown of inconclusive results are - do they have large fractions of spheroid vs point source? Does this support the same claim? --> hard to judge. Its masked by the intrinsic shape of inconclusive fraction space being a weird U.

\subsection{Comparison with HST Morphologies}\label{subsec:galzoocomp}

%to get around the galaxy zoo issue AND the repeat citation of Simmons 2017, replace all the references after the first with a "hereafter, S17"

\begin{figure*}[t!]
\includegraphics[width=1\textwidth,]{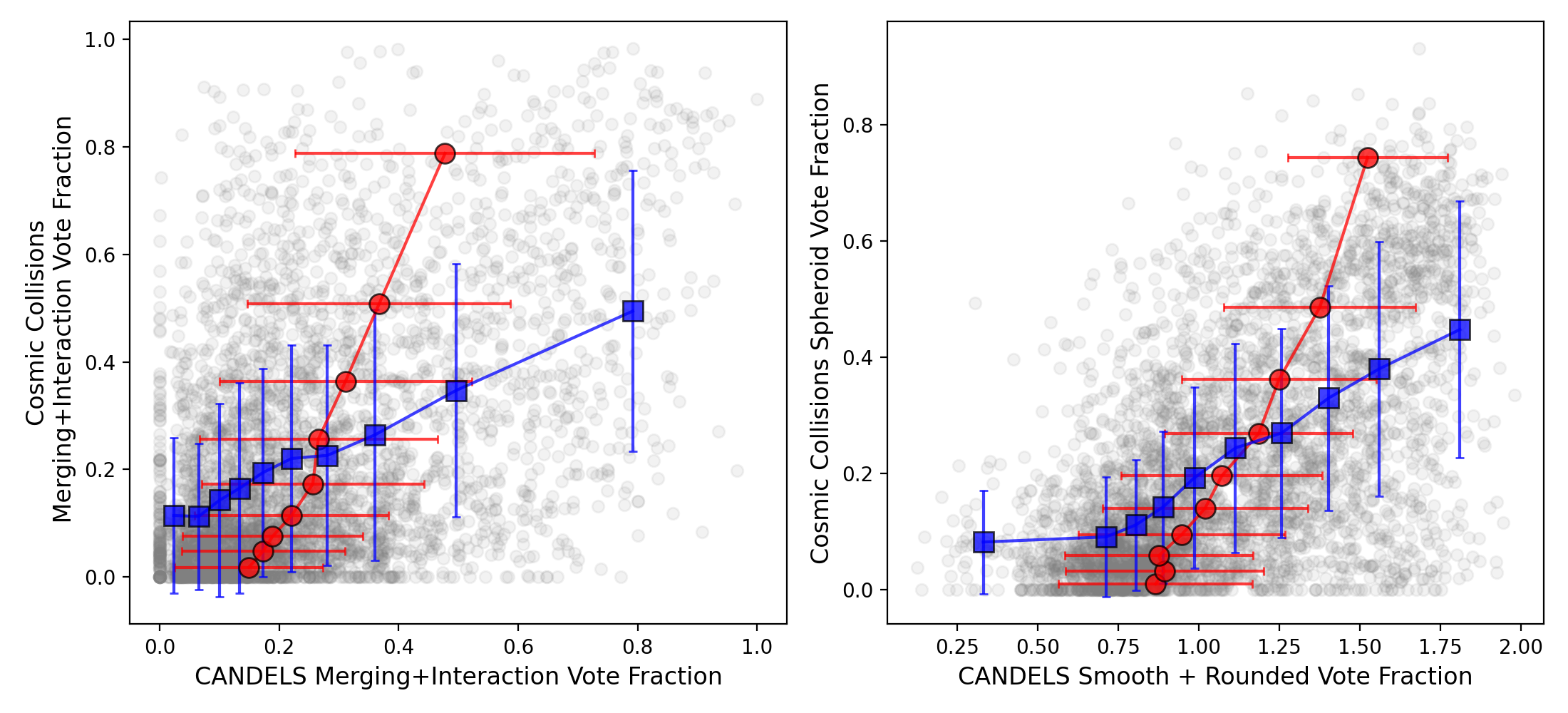}
    \caption{Scatter plots showing the comparison between qualitatively similar combinations of classifications between Cosmic Collisions and those of \cite{galaxyZooCANDELS}. Gray points show the fraction or sum values for individual objects. Blue squares show the average value of Cosmic Collisions classifications binned by 10\% percentiles in the \cite{galaxyZooCANDELS} classifications. Red circles show the average value of \cite{galaxyZooCANDELS} classifications binned by 10\% percentiles in the Cosmic Collisions classifications. Error bars on red and blue points show the region enclosing the middle 68\% of values in that bin. Left panel: the weighted sum of `Merger' and `Interacting Companion(s)' votes in Task 4 of Cosmic Collisions, compared to the weighted sum of `Merging', `Tidal Debris', and `Both' in Task 16 of \cite{galaxyZooCANDELS}. Right panel: the weighted fraction of `Spheroid' votes in Tasks 1 and 2 of Cosmic Collisions, compared to the weighted sum of `Smooth' votes in Task 0 and `Completely round' votes in Task 1 of \cite{galaxyZooCANDELS}.
\label{fig:class_comparison}}
\end{figure*} 

\cite{galaxyZooCANDELS} undertook a large scale visual classification of data from the \textit{HST} CANDELS \citep{CANDELS, koekemoer2011_candels} Survey on Galaxy Zoo which includes observations of the COSMOS and UDS fields. In addition, their analysis includes a direct comparison with \cite{jeyhan2015}, which is the work our classification scheme is based on. There are 4187 galaxies in our sample that are also in the CANDELS sample. We restrict this subsample using subset\_flag$<$50 and q\_flag=0. This left 3984 objects with reasonable visual classifications in both projects.

The classification system in \cite{galaxyZooCANDELS} is quite different from ours. They collect 40 classifications per object instead of our 20, and their workflow has a more complex decision tree \citep[see Figure 1 in][for a detailed description]{galaxyZooCANDELS}. Their workflow consists of 16 Tasks, organized into five Tiers based on the number of times the decision tree branches before reaching a Task. For example, a Task shown for all objects is 1st Tier, and a Task shown after giving three responses that result in alternate paths is 4th Tier. By this definition, all Tasks in Cosmic Collisions are either 1st or 2nd Tier. There are no Tasks which are precisely the same between the two works, and any Task beyond 1st or 2nd Tier in the \cite{galaxyZooCANDELS} decision tree will be contaminated by a large number of objects that never or almost never reached a particular question when classified. This presents multiple challenges for finding one-to-one comparisons between the two samples.

To overcome these difficulties, we examine the relationship between the two combinations of lower tier classifications which describe the most similar qualities. We compare the sum of weighted of `Merger' and `Interacting Companion(s)' fractions in Task 4 of Cosmic Collisions to the sum of weighted `Merging', `Tidal Debris', and `Both' fractions in Task 16 of \cite{galaxyZooCANDELS}. We also compare the weighted fraction of `Spheroid' votes in Tasks 1 and 2 of Cosmic Collisions to the sum of weighted `Smooth' fractions in Task 0 and `Completely round' fractions in Task 1 of \cite{galaxyZooCANDELS}. We restrict the populations being compared based on whether the relevant Task was reached in at least 10\% of classifications that the object received; this mostly serves to eliminate unclassifiable objects. Figure \ref{fig:class_comparison} shows the comparison between the CANDELS and Cosmic Collisions classifications.

In general, when qualitatively similar combinations of classifications are overplotted, they trace one another within uncertainty across most bins, showing broad agreement between the schemes. There is, however, significant scatter in both panels. Much of this scatter can be attributed to the categories not being perfectly matched, and the intrinsic uncertainty of visual classification, but many objects have stark disagreement. In the left panel, these major disagreements are relatively symmetric, i.e. neither work tends more towards finding interactions when the other does not. In the right panel, \cite{galaxyZooCANDELS} appears to tend to find a large number of smooth, rounded objects that Cosmic Collisions does not consider spheroids.

%Soften the language here. The trends broadly agree, except for the highest bins. One possible reason for that is the greater spread of galaxies, but it can also be partially attributed to the discrepancies that we've already explained 

We visually inspected many of the most egregious disagreements and find that most discrepancies are due to improved resolution and sensitivity between HST and JWST, though some differences are attributable to wavelength effects as well. Galaxies which appear to be interacting in \citep{galaxyZooCANDELS} but not in this work tend to be poorly resolved in HST imaging but clearly resolved with JWST; these objects appear to overlap or share a common envelope when viewed with HST, but become distinctly separated in JWST imaging. Likewise, objects which appear smooth and rounded in HST imaging often have shape and structure in JWST imaging, explaining the discrepancies in the right panel of Figure \ref{fig:class_comparison}. 

Galaxies which appear to be non-interacting in HST imaging but show interaction signatures in this work tend to appear faint in HST imaging; their tidal features, irregular structure, and previously undetected neighbors become visible in JWST imaging. In low redshift cases, where HST and our choices of JWST filters are probing the same stellar populations, this is truly a sensitivity issue. However, at $z>2.5$, even the reddest band used to classify HST images (F160W) begins to lose flux past the 4000\r{A} break, with the two bluer filters dropping out well before this, exacerbating these discrepancies .

\subsection{Serendipitous Science}\label{subsec:leyna}

\begin{figure*}[t]
\includegraphics[width=1\textwidth,]{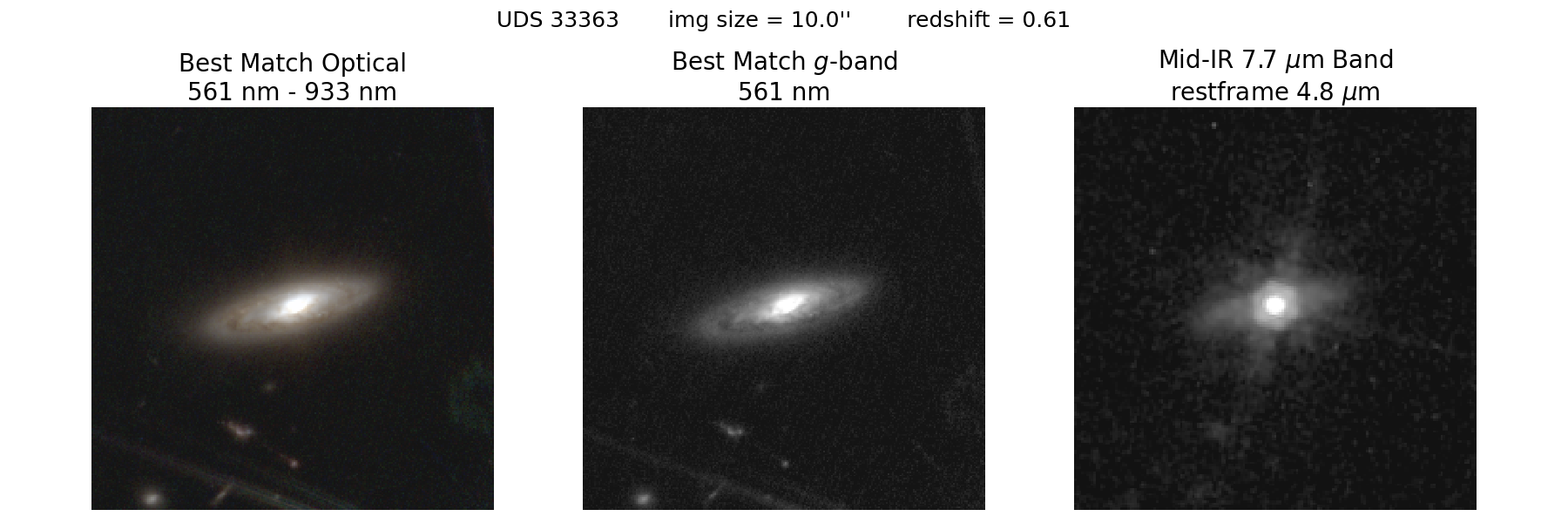}
    \caption{The visual classification image of UDS 33363, part of Season 1. This is an example of an object where only the disk is visible in the rest-optical, but which shows a bright point-source core in the mid-IR. This may be a morphological indicator of an AGN unique to mid-IR surveys.
\label{fig:agn_candidate}}
\end{figure*} 

Citizen science is a useful tool for identifying interesting objects beyond the initial scope of the project. Here, we provide a first look at a possible method of identifying AGN candidates morphologically using combined NIRCam and MIRI imaging, which will be the subject of a forthcoming paper (Bajaj et al.\ in prep). 

In the talk pages, users noted a category of objects which had disk-like morphologies in the rest-frame optical but which appeared to be bright point sources in the mid-IR, often superimposed on top of a mid-IR structure which is comparable to that of the optical. Figure \ref{fig:agn_candidate} gives an example of one such object. 

Following the identification of this category of objects, we searched through our catalog of images and found 63 similar objects. These objects may be obscured AGN, providing a morphological path for identification in future surveys. %These candidates are being investigated further in \cite{Bajajinprep}.

Interestingly, the overwhelming majority of this subsample are morphologically undisturbed, isolated systems, indicating possible secular supermassive black hole growth. If these candidates are confirmed as obscured AGNs, this has implications for the importance of secular evolution in the accretion history of black holes. 

It is important to note the redshift range at which this technique is likely to be effective. Because F770W is relatively close to the wavelength range at which galaxy emission transitions from primarily starlight to primarily warm dust, this filter probes different physical structures across cosmic time. At $z<0.2$, F770W is probing both PAH and the AGN power law; however, PAH emission is unlikely to produce a disproportionately bright core. Between $z\sim0.4$ and $z\sim2$, F770W is well within the stellar minimum, so excess flux is reasonably attributed to an AGN. Above $z\sim3$, F770W is probing the old stellar population. Utilizing redder filters may allow for this technique to be extended to higher redshift, provided the PSF remains small enough to distinguish the core from the galaxy at large. 

\section{Summary and Future Work}\label{sec:conclusions}

We collected and processed morphology classifications by citizen science volunteers for 14,657 mid-IR selected galaxies at cosmic noon. We present weighted and unweighted classification fractions for morphology, interactions, and surface features, as well as a number of useful descriptors to aid usage of our value-added catalog. We find that

\begin{itemize}
    \item Evolution of irregular and interacting populations is largely in line with previous works in terms of evolution with redshift. However, we find a lower interaction rate. 
    \item Galaxies detected in the mid-IR show differences in their morphologies as a function of redshift when compared with optically or near-IR selected samples. In the local universe, they are more disk-like, because we are generally selecting for more dust-rich galaxies; in the distant universe, they are more compact, possibly due to over-representation of AGNs. 
    \item The resolution, sensitivity, and wavelength range of JWST can drastically alter the way a galaxy is classified in comparison to HST. Because the disparity is often due to improved imaging, in cases of extreme difference, JWST morphologies are likely more reliable. 
\end{itemize}

\begin{acknowledgments}
A. Kirkpatrick. and G. Troiani gratefully acknowledge support from JWST-AR-07912 and JWST-GO-01837. 
This work is based on observations made with the NASA/ESA/CSA James Webb Space Telescope. The data were obtained from the Mikulski Archive for Space Telescopes at the Space Telescope Science Institute, which is operated by the Association of Universities for Research in Astronomy, Inc., under NASA contract NAS 5-03127 for JWST. These observations are associated with program ID 1837.
This publication uses data generated via the Zooniverse.org platform, development of which is funded by generous support, including a Global Impact Award from Google, and by a grant from the Alfred P. Sloan Foundation. 
J. S. Dunlop acknowledges the support of the Science and Technology Facilities Council.
P. G. Pérez-González acknowledges support from Spanish Ministerio de Ciencia e Innovación MCIN/AEI/10.13039/501100011033 through grant PGC2018-093499-B-I00.

\end{acknowledgments} 


\vspace{5mm}
\facilities{HST, JWST}


\software{astropy \citep{astropy},   
          Source Extractor \citep{SexTractor},
          Trilogy \citep{trilogy}. ChatGPT \citep{openai2022chatgpt} was used to produce an abstract summary; all LLM output was edited and checked for accuracy by human authors.}

%% Appendix material should be preceded with a single 
%\appendix
%command.



\bibliographystyle{aasjournal}
\bibliography{bibliography}



\end{document}